\documentclass[fleqn,usenatbib]{mnras}

\usepackage{newtxtext,newtxmath}

\usepackage[T1]{fontenc}

\DeclareRobustCommand{\VAN}[3]{#2}
\let\VANthebibliography\thebibliography
\def\thebibliography{\DeclareRobustCommand{\VAN}[3]{##3}\VANthebibliography}

\usepackage{graphicx}	
\usepackage{newtxtext,newtxmath}
\usepackage{amsfonts}
\usepackage{longtable}
\usepackage{CJKutf8}
\usepackage{multicol,multirow}
\usepackage{orcidlink}

\newcommand{\dm}{\,pc cm$^{-3}$}

\title[ASKAP FRB Pipeline Algorithms]{Systematic performance of the ASKAP Fast Radio Burst search algorithm}

\author[H. Qiu et al.]{Hao Qiu
(\begin{CJK*}{UTF8}{gbsn}邱昊\end{CJK*})\orcidlink{0000-0002-9586-7904}$^{1}$\thanks{E-mail: hao.qiu@skao.int},
Evan F. Keane\orcidlink{0000-0002-4553-655X}$^{2}$,
Keith W. Bannister\orcidlink{0000-0003-2149-0363}$^{3}$,
Clancy W. James\orcidlink{0000-0002-6437-6176}$^{4}$
\& Ryan M. Shannon\orcidlink{0000-0002-7285-6348}$^{5}$
\\
$^{1}$ SKA Observatory, Jodrell Bank, Lower Withington,
Macclesfield, SK11 9FT, UK\\
$^{2}$School of Physics, Trinity College Dublin, College Green, Dublin 2, Ireland\\
$^{3}$Australia Telescope National Facility, CSIRO Space \& Astronomy, P.O. Box 76, Epping, NSW 1710, Australia\\
$^{4}$International Centre for Radio Astronomy Research, Curtin Institute of Radio Astronomy, Curtin University, WA 6845, Australia\\
$^{5}$Centre for Astrophysics and Supercomputing, Swinburne University of Technology, Hawthorn,VIC 3122, Australia
}

\date{Accepted XXX. Received YYY; in original form ZZZ}

\pubyear{2021}

\begin{document}
\label{firstpage}
\pagerange{\pageref{firstpage}--\pageref{lastpage}}
\maketitle

\begin{abstract}
Detecting fast radio bursts (FRBs) requires software pipelines to search for dispersed single pulses of emission in radio telescope data.
In order to enable an unbiased estimation of the underlying FRB population,
it is important to understand the algorithm efficiency with respect to the search parameter space and thus the survey completeness.
The Fast Real-time Engine for Dedispersing Amplitudes
({\sc{fredda}}) search pipeline is a single pulse detection pipeline designed to identify radio pulses over a large range of dispersion measures (DM) with low latency. It is used on the Australian Square Kilometre Array Pathfinder (ASKAP) for the Commensal Real-time ASKAP Fast Transients (CRAFT) project .
We utilise simulated single pulses in the low- and high-frequency observation bands of ASKAP to analyse the performance of the pipeline and infer the underlying FRB population.
The simulation explores the Signal-to-Noise Ratio (S/N) recovery as a function of DM and the temporal duration of FRB pulses in comparison to injected values.
The effects of intra-channel broadening caused by dispersion are also carefully studied in this work using control datasets.
Our results show that for Gaussian-like single pulses, $> 85 \%$ of the injected signal is recovered by pipelines such as {\sc fredda} at DM < 3000 \dm using standard boxcar filters compared to an ideal incoherent dedispersion match filter. 
Further calculations with sensitivity implies at least $\sim 10\%$ of FRBs in a Euclidean universe at target sensitivity will be missed by {\sc fredda} and {\sc heimdall}, another common pipeline, in ideal radio environments at 1.1 GHz.

\end{abstract}

\begin{keywords}
fast radio bursts -- methods: data analysis -- software: simulations
\end{keywords}



\section{Introduction}

Fast Radio Bursts \citep[FRBs;][]{2007Sci...318..777L} are dispersed bright single pulses of extragalactic origin \citep[][]{2017Natur.541...58C,2019Sci...365..565B}. Like pulses observed from pulsars, FRBs are subject to propagation effects: for FRBs this arises when passing through the extragalactic and interstellar media. Their most distinctive feature is a frequency-dependent time delay caused by dispersion when travelling through cold ionised media. The magnitude of this effect is given by the integral of the free electron column density along the line of sight, known as the dispersion measure (DM). A key feature of FRBs, and perhaps their working observational definition~\citep{2013Sci...341...53T,2016MNRAS.459.1360K}, is that their DMs exceed the maximum contribution from the Milky Way, indicating an extragalactic origin. In support of this, there is an strong observed correlation between the extragalactic DM and luminosity distance~\citep{2020Natur.581..391M}. Additional propagation effects such as scatter broadening and Faraday rotation caused by turbulent or clumping plasma structures and magnetic fields along the line of sight can have a significant effect on the pulse morphology of FRBs \citep{2013ApJ...776..125M,2019Sci...366..231P,2021ApJ...922..173C}. These propagation effects have shown to be useful probes of the extragalactic medium \citep{2020ApJ...901..134S} and, for example, can be used as an independent measurement to reveal the density and magnetic fields of foreground galaxy haloes \citep{2019Sci...366..231P}.

The search for FRBs is performed using single pulse search pipelines for both real-time observations and post-facto offline processing; typically there is a trade-off between processing speed and the degree of thoroughness in terms of data cleaning~\citep{2018MNRAS.473..116K}. The core algorithm of single pulse detection is dedispersing the data to a large number of trial DM values, and then identifying high signal-to-noise ratio (S/N) pulses within these data for a range of pulse durations.
Many algorithms have been used to detect FRBs, including but not limited to: 
{\sc seek} \citep{2007Sci...318..777L}, {\sc destroy} \citep{2012MNRAS.425L..71K}, {\sc dedisperse-all} \citep{2014ApJ...792...19B}, {\sc presto} \citep{2016Natur.531..202S}, 
{\sc heimdall} \citep{Barsdell2012}, 
{\sc bonsai} \citep{2018ApJ...863...48C}, {\sc fredda} \citep{2017ApJ...841L..12B,2019ascl.soft06003B}, 
{\sc amber} \citep{amber} and {\sc astroaccelerate} \citep{armour2011gpu,astroaccel,adamek2020single}. 
For the typical case where rapid follow-up is needed, it is the real-time pipelines that make the discoveries. These real-time pipelines are often optimised for low latency while maintaining as complete a search sensitivity as affordable.

The ideal pipeline should have a stable uniform characterized performance across the parameter space during interference-free observations.
The characteristic response of the algorithm or pipeline, across its searched parameter space, impacts how one interprets the output results. 
Pipeline performance can be understood through the analysis of detection rates and also the S/N reported by the algorithm.
Incomplete signal recovery of pulses and other systematic biases during single pulse searches will cause lower S/N and fewer pulses to be detected above the threshold; these systematic errors are often related to the DM and pulse width of the pulse. 
The underlying population distribution estimated from the observed sample is crucial to cosmological parameter estimations using FRBs \citep{2019MNRAS.487.5753C,2020MNRAS.494..665L,2022MNRAS.509.4775J}.
This paper focuses on the identification of this S/N response function by using simulated pulses injected in observational format data for two FRB search pipelines:
{\sc fredda} and {\sc heimdall}.

Mock FRB injections systems have been deployed in major FRB observing facilities such as CHIME \citep{2018ApJ...863...48C}, 
UTMOST \citep{2019MNRAS.488.2989F} and the GBT \citep{2020MNRAS.497..352A}.
The {\sc greenburst} system for GBT developed in \citet{2020MNRAS.497..352A} measured the system recall curve, measuring a 100\% recovery of injected bursts at $\mathrm{S/N}\gtrsim 12$.
Recent systematic injection tests by \citet{2021MNRAS.501.2316G} showed that {\sc heimdall} used on UTMOST was able to recover over 90\% of the synthetic injections above a S/N threshold of 9.

The Fast Real-time Engine for Dedispersing Amplitudes \citep[{\sc fredda};][]{2017ApJ...841L..12B,2019ascl.soft06003B}
is the search pipeline used on the Australian Square Kilometre Array Pathfinder (ASKAP) for the Commensal Real-time ASKAP Fast Transients (CRAFT) project.  
It is a GPU-based implementation of the Fast Dispersion Measure Transform \citep[FDMT;][]{2017ApJ...835...11Z}, a rapid dedispersion trial algorithm to detect dispersed single pulses.
{\sc fredda} aims to detect FRBs from the incoherent data stream of ASKAP in low latency to trigger the download of baseband ring buffer data for interferometry.


In this paper, we examine the performance of {\sc fredda} using 
dispersed pulses simulated over a large range of dispersion measures and pulse widths.
The aim is to understand the signal recovery fraction from the detection pipeline to infer the real sensitivity threshold of ASKAP FRBs and other major FRB surveys.
We use {\sc heimdall}, a widely-used effective search pipeline, to perform a comparison search on the simulated data. This allows us to verify the simulated bursts and compare the S/N algorithms between these two software. 
The simulation methods are described in \S~\ref{sec:model}.
The data processing setup and results are presented in \S~\ref{sec:data}. We then, in \S~\ref{sec:analysis}, analyse and interpret these results, discussing possible implications for how to interpret the observed FRB samples.

\section{Modelling}

\label{sec:model}
\subsection{Simulation data format}

For most detection pipelines acting in a blind survey mode, the input data are the Stokes I dynamic spectra, i.e. \textsc{sigproc} filterbank files\footnote{\texttt{https://sigproc.sourceforge.net/}}.
The data are usually stored with millisecond time resolution with no coherent dedispersion applied. As a result, the microsecond fine structure seen in some FRBs (e.g \citealt{2018MNRAS.478.1209F,2020MNRAS.497.3335D,2018Natur.553..182M,nimmo2020}) cannot be resolved in the initial detection. If a pulse is detected with low latency and a higher resolution data product is in temporary storage then analysis at higher resolution is possible post-facto.

In this work, we use single Gaussian pulses to mimic the smeared pulse appearance of most observed FRBs~\citep{2021ApJ...923....1P}.
The simulated pulses have a constant injected S/N of $50$. The data are generated at time and frequency resolution of $0.1$~ms and $0.1$~MHz respectively. The data are then down-scaled by a factor of $10$ in both dimensions. This is done so as to introduce various smearing effects and create a precise pulse profile. The reduced resolutions aim to match closely those of ASKAP data (see Table~\ref{tab:singlepulse}). The pulses are injected into a background of Gaussian white noise. Together these steps simulate ideal observation conditions.

\subsection{Simulation format}
\begin{table}
\caption{Simulated Dataset properties}
\begin{tabular}{l|l|l}
\hline
Dataset &Gaussian & Scattering\\
\hline
     DM (\dm)& 0--3000 & 0--3000 \\
     DM step &50&500\\
     Intrinsic Width (ms)& 0.5--11.0 &  1 \& 5\\
     $\sigma_{\rm intrinsic}$ step (ms) & 0.5 &--\\
     Scattering time $\tau_{\rm{scat}}$ (ms)&0 & 0.5--10\\
     $\tau_{\rm{scat}}$ step&--&0.5\\
     \hline
     $\rm{N{bits}}$ & \multicolumn{2}{c}{8}\\
     $\rm{N{chan}}$ & \multicolumn{2}{c}{336}\\
     $\rm{t_{samp}}$ (ms)& \multicolumn{2}{c}{1} \\
     $\Delta\nu$ (ms)& \multicolumn{2}{c}{1} \\
     High Freq Band (GHz)& \multicolumn{2}{c}{1.1--1.436} \\ 
     Low Freq Band (GHz)& \multicolumn{2}{c}{0.764--1.1} \\
     Injected S/N &\multicolumn{2}{c}{50} \\
     Number of pulses & \multicolumn{2}{c}{50}\\
\hline
\end{tabular}
    \label{tab:singlepulse}
\vspace{-10.0pt}
\end{table}

The data recorded for ASKAP FRB searches are also filterbanks. ASKAP observes at radio frequencies between 0.7 and 1.8 GHz. The ASKAP beamformers create $36$ Stokes I beams per dish, i.e. $36\times N_{\rm dish}$ data streams~\citep{2014JAI.....350004C}. The CRAFT backend adds these incoherently to create $36\times 1$ data streams.
The CRAFT pipeline is designed to search such filterbank data generated 
with a time resolution typical between 0.7-1.7 ms and with a bandwidth of 336 MHz. In this work we use the two standard frequency bands employed by CRAFT (see Table~\ref{tab:singlepulse}). The ASKAP data streams
are typically channelized to $336\times 1\mathrm{MHz}$ channels with a time resolution of 1.26~ms (before 2019, later changed to 1.73 ms for lower frequency band searches). The resolution of the simulated dynamic spectrum in this work is intended to be similar to the resolution of the current incoherent sum data stream during CRAFT observations. The format is $336\times 1$ MHz channels and a time resolution of 1 ms recorded in 8-bit data(see Table~\ref{tab:singlepulse}).

\subsection{Model of injected pulses}
\begin{figure}
    \centering
    \includegraphics[width=0.95\columnwidth]{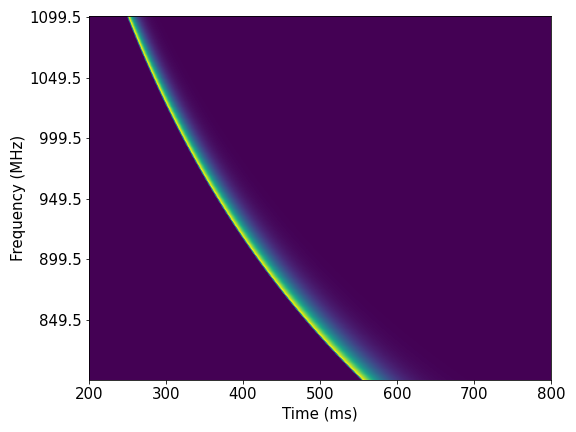}
    \caption[Simulation of dispersed Gaussian pulse]{A simulated Gaussian pulse with an intrinsic width of $\sigma$=0.5 ms, a dispersion time delay of $\rm{DM}=100$ \dm, a scattering time of $\tau_{1\rm{GHz}}=10$ ms, scattering index $\alpha=4$ and flat spectral index. The time resolution of the filterbank is 1 ms, with no background noise. The slight variability in the intensity of the pulse shows the pulse energy distributed over different number of time samples in each channel caused by intra-channel dispersion and different pulse arrival time.}
    \label{fig:basics dispersion sweep}
\end{figure}

We assume a single Gaussian as the underlying profile of each pulse, i.e. in $0.1$-MHz $0.1$-ms resolution with the following equation. For channel $i$ the flux of the pulse is:
\begin{equation}
S_i(t)=\frac{A}{\sqrt{2\pi\sigma_i^2}} \exp \left[\frac{-(t-t_0-t_{\rm DM}(\nu_{\rm i})^2}{2\sigma_i^2} \right],
\label{eq:gaus_ch2}
\end{equation}
where $A$ is an amplitude scale factor, $t_0$ is the time reference of the burst at the reference frequency which we take to be the top of the band, i.e. the centre of the highest frequency channel. The standard deviation of the intrinsic Gaussian is $\sigma_{\rm i}$ and  
the dispersion delay time, $t_{\rm DM}(\nu_{\rm i})$ relative to the reference frequency and is proportional to the DM according to:
\begin{equation}
t_{\rm DM}(\nu_{\rm i})=4.15\ \mathrm{DM}(\nu_{\mathrm{top}}^{-2}-\nu_i^{-2})\;\mathrm{ms},
\label{eq:tidm_ch2}
\end{equation}
where $\nu_{\rm i}$ is the centre frequency of the i$^{\rm th}$ channel and $\nu_{\rm{top}}$ is the centre frequency of the top channel. 
The intra-channel dispersion smearing is the dispersion delay time within one channel, causing the pulse to be broadened within the channel~\citep{2013ApJS..205....4C}.
The dispersion smearing in the i$^{\rm th}$ channel is:
\begin{equation}
\Delta t_{\rm DM}=(8.3\times 10^{-3}  )\ \mathrm{\Delta\nu\ {DM}}\  \nu_i^{-3} \rm{ms}.
\label{eq:smear_ch2}
\end{equation}
Here $\Delta\nu$ is the channel bandwidth in units of MHz and $\nu_{\rm i}$ is the channel frequency in GHz. We interpret $\Delta t_{\rm DM}$ as the full-width half maximum (FWHM) of a Gaussian with width $\sigma_{\rm DM}=\Delta t_{\rm DM}/(2\sqrt{2\ln 2})$. 
For our simulated pulses we define the standard deviation in Eq. \ref{eq:gaus_ch2} as the quadrature sum: 
\begin{equation}
\sigma_i=(\sigma_{\rm{intrinsic}}^2+\sigma_{\mathrm{DM}}^2)^{1/2}.
\label{eq:smearing}
\end{equation}
This generates a Gaussian pulse profile with a typically very small amount of dispersion smearing across the $0.1$~MHz channels. Then reducing the resolution in both frequency and time by a factor of $10$ in each produces dispersion smearing in each channel that is $10$ times higher accounting for the majority of the effect. An example of a resultant pulse is shown in Figure \ref{fig:basics dispersion sweep}.

For this simulation the software we have developed allows for the addition of scatter broadening to the pulse profile. This is achieved by the convolution with a one-sided frequency-dependent exponential decay function.
The scattering index is set as $\alpha = 4$ based on measurements from scattering in pulsar observations \citep{bhat2004}:
\begin{equation}
S_i(t)=\begin{cases}
    \exp\left[-\frac{(t-t_p)}{\tau({\nu_i}/{{1\rm{GHz}}})^{-\alpha}}\right], & (t\geq t_p),\\
    0 ,         & (t< t_p).
\end{cases}
\label{eq:scat}
\end{equation}
where $1$~GHz is the reference frequency and $\rm{t_p= t_0+t_{DM}(\nu_i)}$.

\subsection{S/N rescaling}
\label{sec:rescaling}

The pulse is independently modelled in each channel based on equation \ref{eq:gaus_ch2}. 
Down-sampling causes a DM smearing effect on the pulse profile, broadening its width.
The S/N of an injected pulse is calculated using a match filter on the dedispersed time series in units of the standard deviation of the noise background. All of the simulation steps are done using 32-bit floats; the very last step in our pipeline is to write out 8-bit files (again to match the ASKAP output). 
This calculation standardizes the S/N of the simulated burst so that we can compare with the S/N measured from the pipelines.
We define the match-filter S/N in the time series as:
\begin{equation}
    \rm{S/N_{filter}=\sqrt{\sum S_i^2}},
    \label{eq:snr_match}
\end{equation}
where $\rm{S_i}$ is the signal intensity of each sample in units of the standard deviation. 

\begin{figure}
\includegraphics[width=0.95\columnwidth]{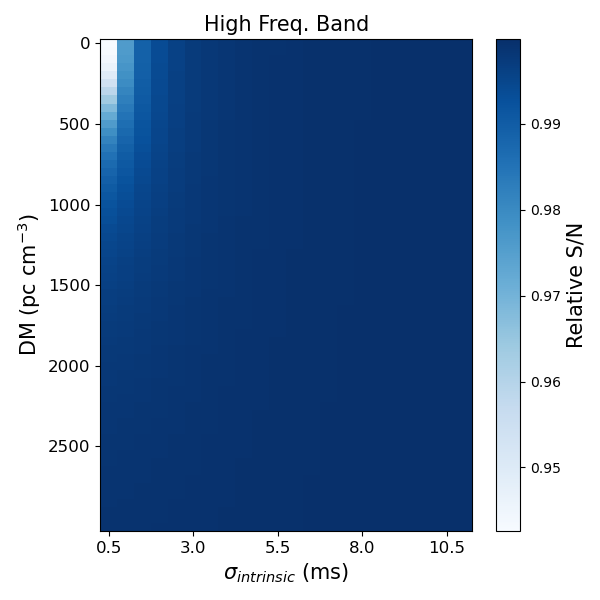}
    \caption{
    The theoretical S/N scaling after resolution downscaling measured by boxcar matchfilters for pulses in the high frequency band.}
    \label{fig:downscale}   
\end{figure}

After being time averaged, the pulse fluence (in units of $\mathrm{Jy\;ms}$) remains constant, but S/N is not conserved. This is shown in Figure \ref{fig:downscale}:
for narrow Gaussian pulses with FWHM less than 1 coarse time sample (the final data time resolution, here $1$~ms), the S/N decreases with burst width. 
The pulse S/N is again re-adjusted, by scaling, to the intended value so that all injected pulses have consistent S/N for the purpose of searching.
Gaussian background noise with a standard deviation of 1 is then added to the array and the data are written to disk in {\sc sigproc} filterbank format. 
Using the match filter equation (Eq.\ref{eq:snr_match}), it is known that the final readout S/N of the pulse under the added noise background will have an approximate uncertainty of 1 standard deviation unit, our output results also match this result.

The bursts are chosen to be scaled to S/N=50 during this process. The response of the algorithm is independent of this choice.
This exact value is arbitrary, we choose to inject bright bursts to guarantee we do not miss candidates below the minimum S/N threshold, as would occur if using injections at S/N=10. (See \S~\ref{sec:snr})


\subsection{Dataset setup}
The parameters describing the datasets are shown in Table \ref{tab:singlepulse}.
The experiment is conducted using two frequency bands that are commonly observed by ASKAP. We refer to the two frequency bands in this work as the high frequency band: 1100 MHz -- 1436 MHz and the low frequency band: 764 MHz -- 1100 MHz. For each band, we generated Gaussian single pulses over a large range in DM and intrinsic pulse width. For each sample in DM and width space, $50$ pulses were injected to effect $50$ trial iterations to sample the noise.
Scattered pulses were created as a separate dataset for additional comparison studies (see \S~\ref{sec:scattering}).

An incoherently dedispersed `zero-DM' dataset was also created as a control sample for the Gaussian pulses.
The pulses in this dataset are the exact same pulses as the original dataset except that the pulse time is aligned across the frequency band.
This dataset was created to avoid the effects of incorrect dedispersion that is unavoidable for blind searches with finite numbers of DM trials. Thus it examined
only the basic boxcar S/N retrieval efficiency of the pipelines.

\section{Pipeline setup}

\label{sec:data}

\subsection{Pipeline Settings}

For {\sc fredda}, we use a basic setup only defining two parameters beyond default values. These are the block size of the data stream to ingest at any one time, and the number of DM trials (these are the {\it -t} and {\it -d} input flags for \textsc{fredda}). We use a block size of $8192$ ($16384$) samples for the high (low) frequency bands to ensure that the widest pulses at the highest DMs are not split across blocks. By default \textsc{fredda} is configured so that boxcar widths searches range from $1$ up to a maximum of 32 samples in steps of $1$ \citep{2017ApJ...841L..12B}.

For {\sc heimdall} we also use mostly default parameters. We set the baseline length parameter ({\it -baseline\_length} flag) to be $20$~s, $10$ times the default as we would need to do when observing a strong pulsar, so as to avoid incorrect statistical estimates for the noise. We further turned down a setting to the friends-of-friends algorithm (the {\it -cand\_sep} flag) to ensure adjacent pulses did not get erroneously associated. In the case of our zero-DM pulses, we remove (using the {\it -rfi\_no\_broad} flag) the usual zero-DM filter~\citep{2009MNRAS.395..410E}. Finally we set the DM range and set the maximum boxcar width to 32 samples, however in the case of {\sc heimdall} this parameter is searched in logarithmic steps of a factor of $2$.

One parameter for \textsc{heimdall} can drastically change its output response; this is the DM tolerance parameter ({\it -dm\_tol} flag). This parameter defines the DM step size, the lower the tolerance the smaller the step size. A range of tolerances are used at different observatories (see Table~\ref{tab:dmtol_surveys}); the default value is $1.25$. \textsc{fredda} uses a constant (frequency-dependent) DM step size for the full DM range, and this feature is not configurable. We perform analyses using $1.25$ and $1.01$ tolerances. The former represents a typical value for searches which have discovered FRBs; the latter is in some sense a fairer comparison to \textsc{fredda}. We show results for a DM tolerance of $1.01$ for \textsc{heimdall}. Coarser tolerance results are worse and can be examined in the supplementary online material for this paper.

\begin{table}
    \centering
    \begin{tabular}{l|c|l}
    \hline
    Survey/Telescope & dm\_tol & Reference\\
    \hline
        Parkes-SUPERB & 1.20 & \citet{2018MNRAS.473..116K}\\
        UTMOST & 1.20 & \citet{2021MNRAS.501.2316G}\\
        DSA-10 & 1.15 & \citet{2019MNRAS.489..919K}\\
        STARE2 & 1.25 & \citet{2020PASP..132c4202B}\\
        Sardinia (Targeted)& 1.01 &\citet{2020ApJ...896L..40P}\\
    \hline
    \end{tabular}
    \caption{Example {\sc heimdall} set up for the {\it -dm\_tol} input flag parameter in different real-time blind surveys and targeted searches}
    \label{tab:dmtol_surveys}
\end{table}

\subsection{Known boxcar efficiency and scallop responses}
\label{sec:matchfilter}

To measure the dispersion of the candidates,
{\sc fredda} uses the fast dispersion measure transform (FDMT) algorithm \citep{2017ApJ...835...11Z}, while {\sc heimdall} uses a brute force dedispersion tree.
In order to understand the performance of the pipeline, we investigate the single pulse search algorithms in isolation from the dedispersion algorithms. 

Both pipelines use boxcar filters. We first calculate the theoretical response on single channel boxcar pulses, where the search filter perfectly matches the shape of the signal. 

\begin{figure}
\centering
\includegraphics[trim={0 0.0cm 0 0.0cm },clip,width=0.8\columnwidth]{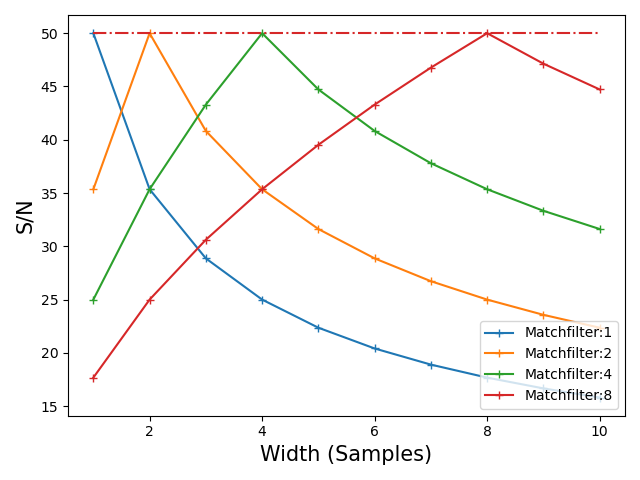}
    \caption{Theoretical S/N recovery of S/N=50 boxcar signals using boxcar match filters with a width of 1, 2, 3, 4 and 8 samples. }
    \label{fig:theoretical_boxcars}
\end{figure}

For the results shown here we ensured that the boxcars were aligned in phase with the time samples, but verified that the S/N fall-off when pulses are out-of-phase with the sampling is as expected. We use filter templates with widths of 1, 2, 4, 8 samples to measure the S/N of boxcar signals (injected S/N of $50$, no noise) with boxcar width between $1-10$~ms. The result in Figure \ref{fig:theoretical_boxcars} shows how using a set of fixed-width match filters, the S/N between the exact filter widths dips. The response is sharply peaked at perfect recovery; the falls off are $(W/B)^{-1/2}$ ($(B/W)^{-1/2}$) when the boxcar width $B$ is greater (less) than the injected pulse width $W$. The overall response shape taken from the maximum achivable S/N drops between matching widths and is commonly known as `scalloping'.

\begin{figure}
\centering
\includegraphics[trim={0 0.0cm 0 0.0cm },clip,width=0.8\columnwidth]{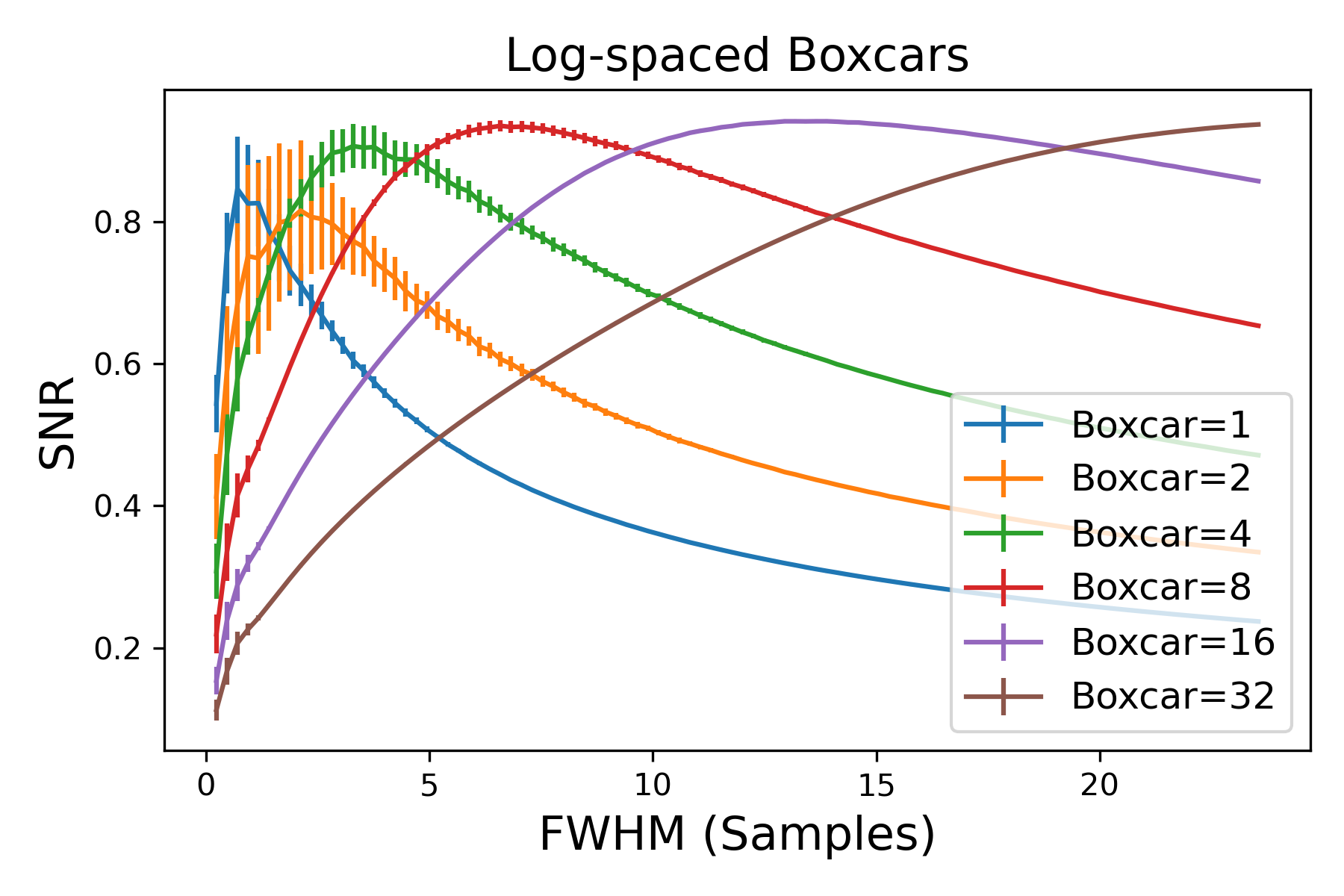}
\includegraphics[trim={0 0.0cm 0 0.0cm },clip,width=0.8\columnwidth]{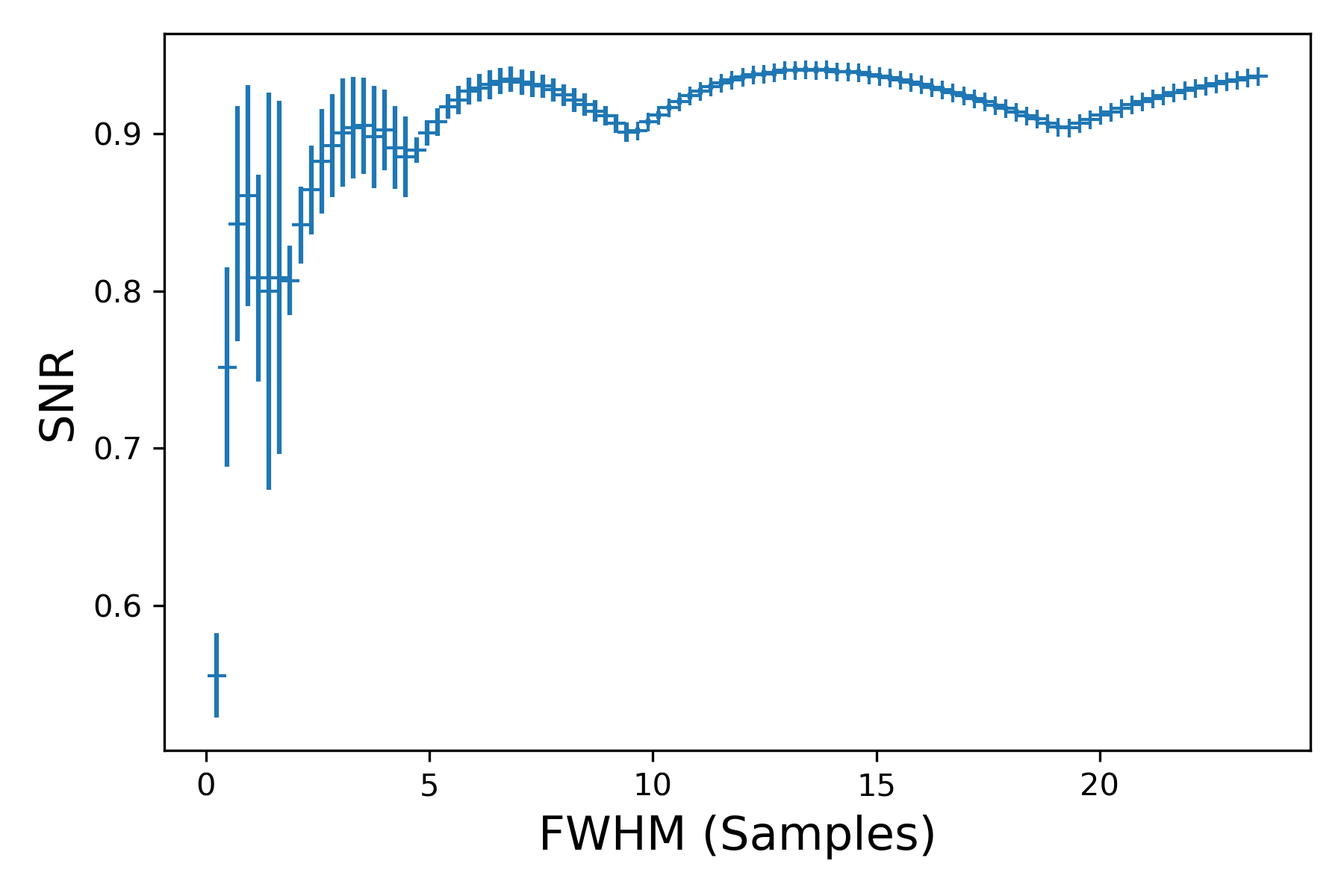}
    \caption{The match filter S/N of Gaussian pulses in 1-ms time resolution using Log-spaced boxcar widths. The upper figure shows the response of each respective match filter while the lower figure shows the resulting response curve by taking the maxima at each width.}
    \label{fig:heimdall_response}
\end{figure}

\begin{figure}
\centering
\includegraphics[trim={0 0.0cm 0 0.0cm },clip,width=0.8\columnwidth]{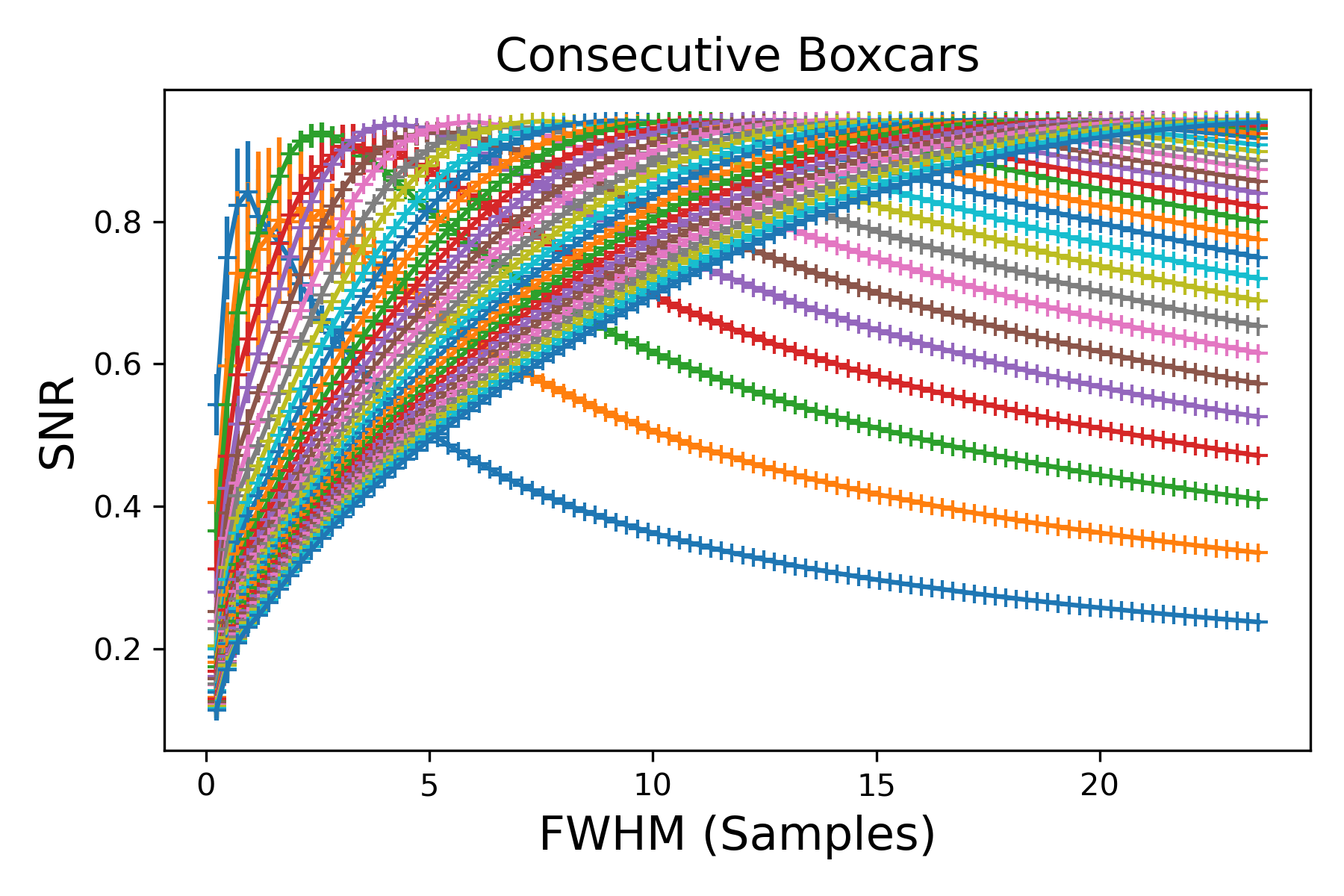}
\includegraphics[trim={0 0.0cm 0 0.0cm },clip,width=0.8\columnwidth]{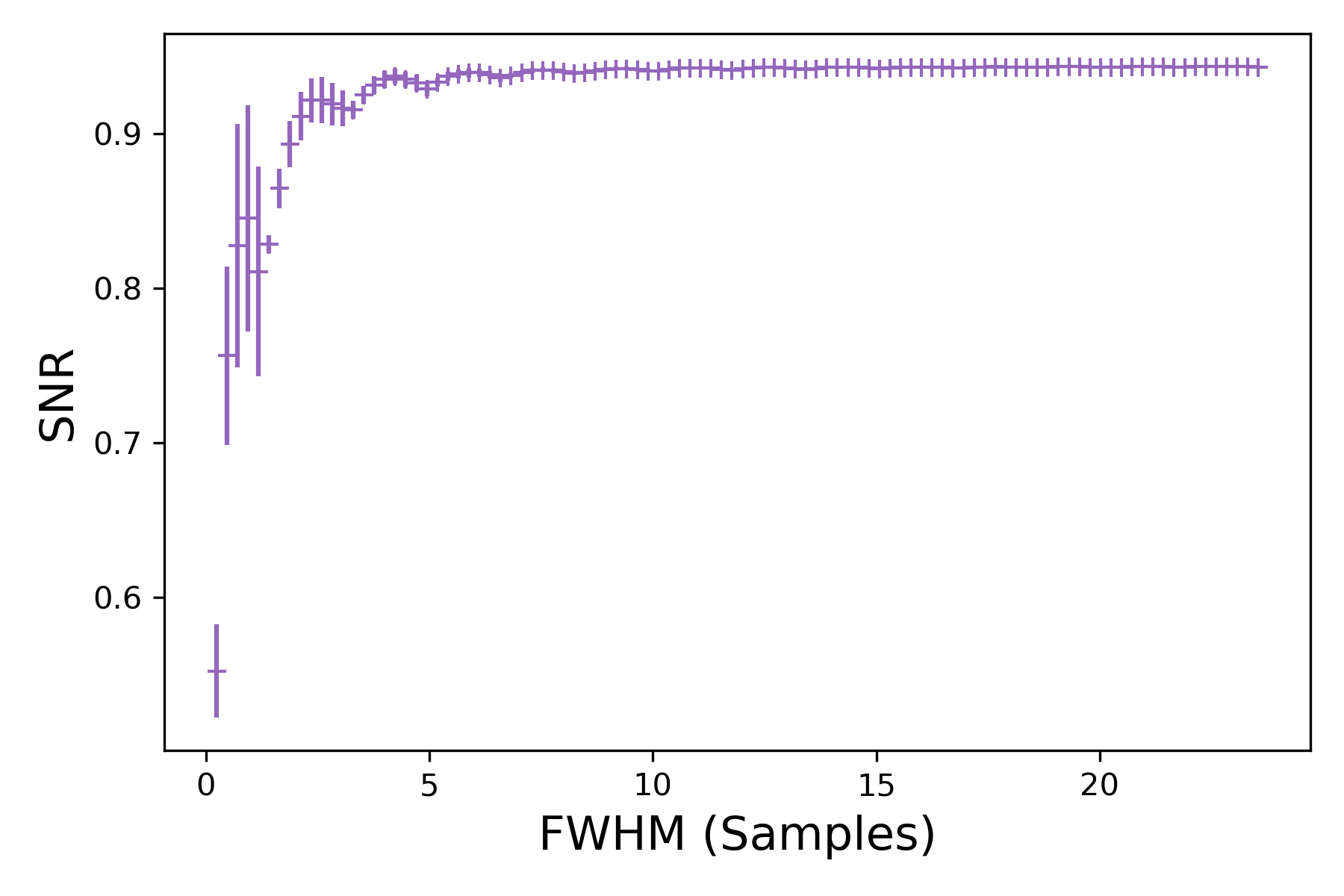}
    \caption{Match filter S/N of Gaussian pulses in 1-ms time resolution using linearly spaced boxcar filters in comparison to Figure \ref{fig:heimdall_response}. By using a set of uniformly spaced and consecutive width boxcar match filters up to 32 samples, it is possible to achieve a more uniform response towards wider pulses.}
    \label{fig:fredda_response}
\end{figure}

The response is different when we inject Gaussian pulses but again
use the boxcar filters of the same widths, as deployed by {\sc heimdall} and {\sc fredda}. The peak response is not perfect but reaches the theoretical maximum of $94.3\%$ recovery~\citep{2015MNRAS.447.2852K,2020MNRAS.497.4654M}, with a smoother falloff than for boxcar pulses. Using a set of boxcar filters with a power of 2 step size, we can observe that {\sc heimdall} suffers between boxcar widths from the scalloping in Figure \ref{fig:heimdall_response}. Due to the low time resolution of the data ingested by the CRAFT pipeline, {\sc fredda} is designed to search consecutive sample widths as this gives a more optimal smooth response curve for wider pulses as shown in Figure \ref{fig:fredda_response}.

It should be noted that both {\sc fredda} and {\sc heimdall} are specifically designed to lower the computation costs and latency when searching for pulses across a large range of width. For Parkes radio telescope data, where {\sc heimdall} was first applied, the time resolution could be as high as 64 $\mu$s. Hence a wider boxcar width range is needed to search for wide pulses and {\sc heimdall} enables a faster search across the range of DM and width in real-time for these observations.
On the other hand, {\sc fredda} was designed specifically for ASKAP data at coarse time resolution, the FDMT algorithm also optimised processing for a large ranges of dedispersion trials.

\section{Results}
\label{sec:analysis}

\subsection{Reported properties of simulated single pulses}
\begin{figure*}
\centering
\includegraphics[trim={0 0.0cm 0 0.0cm },clip,width=0.8\columnwidth]{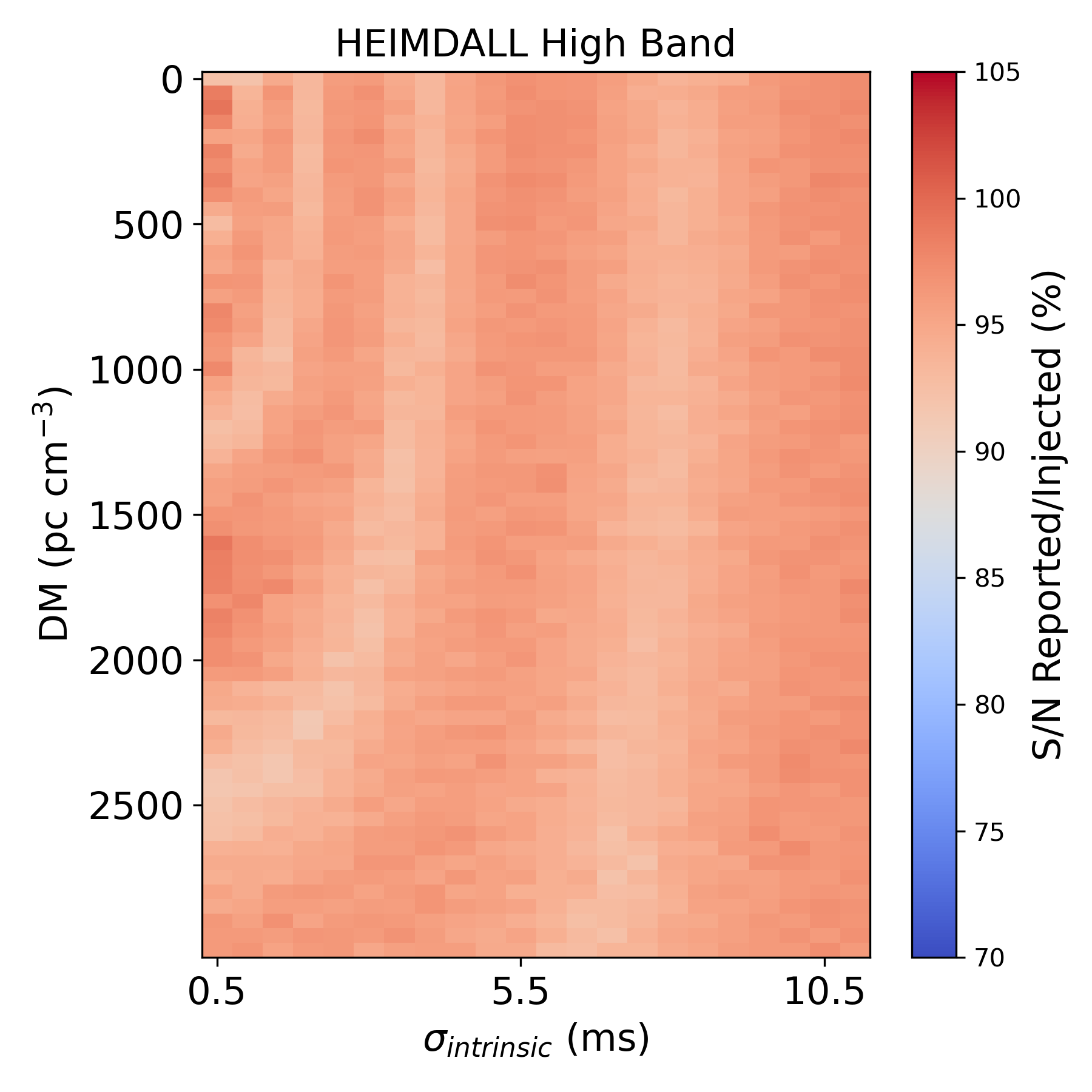}
\includegraphics[trim={0 0.0cm 0 0.0cm },clip,width=0.8\columnwidth]{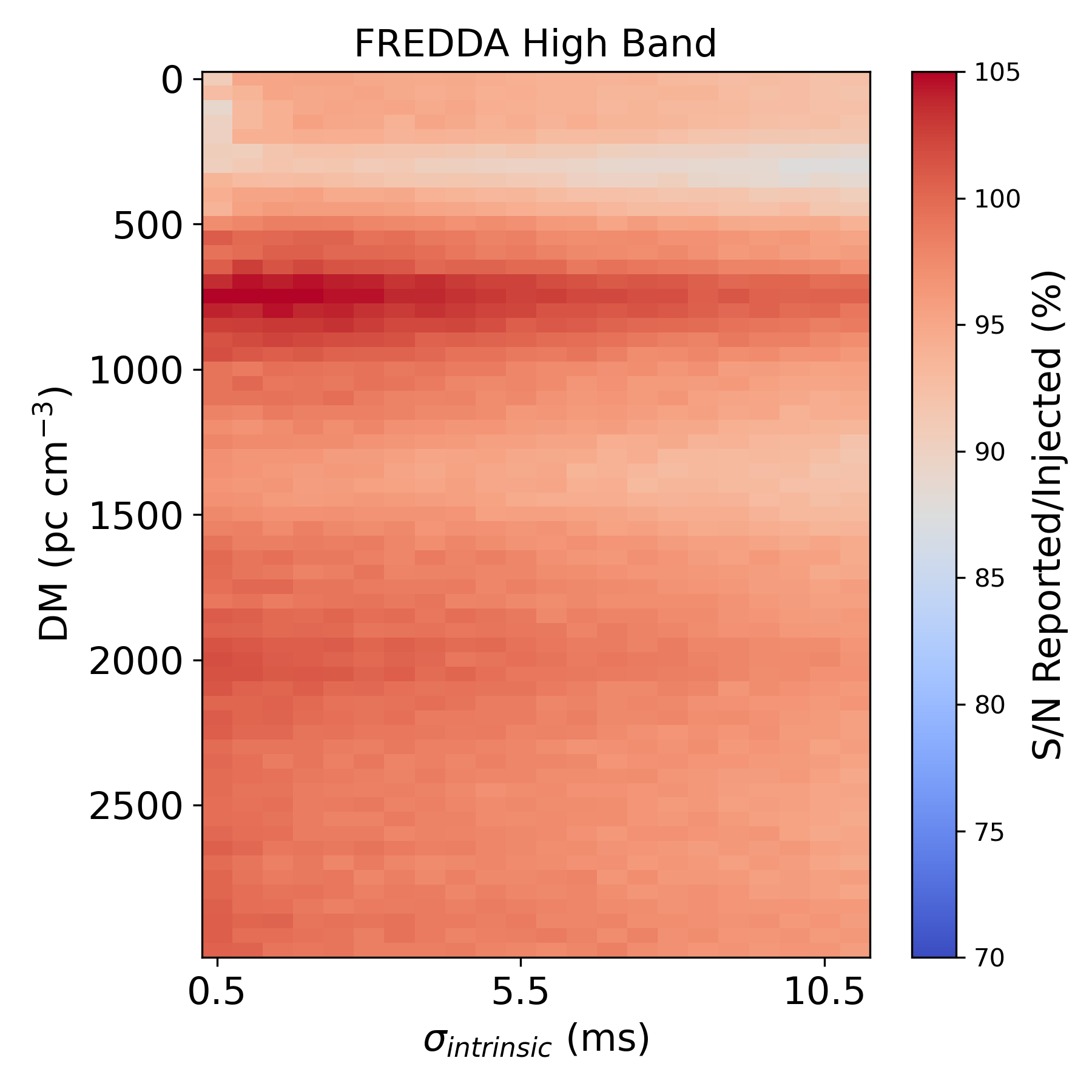}
\\
\includegraphics[trim={0 0.0cm 0 0.0cm },clip,width=0.8\columnwidth]{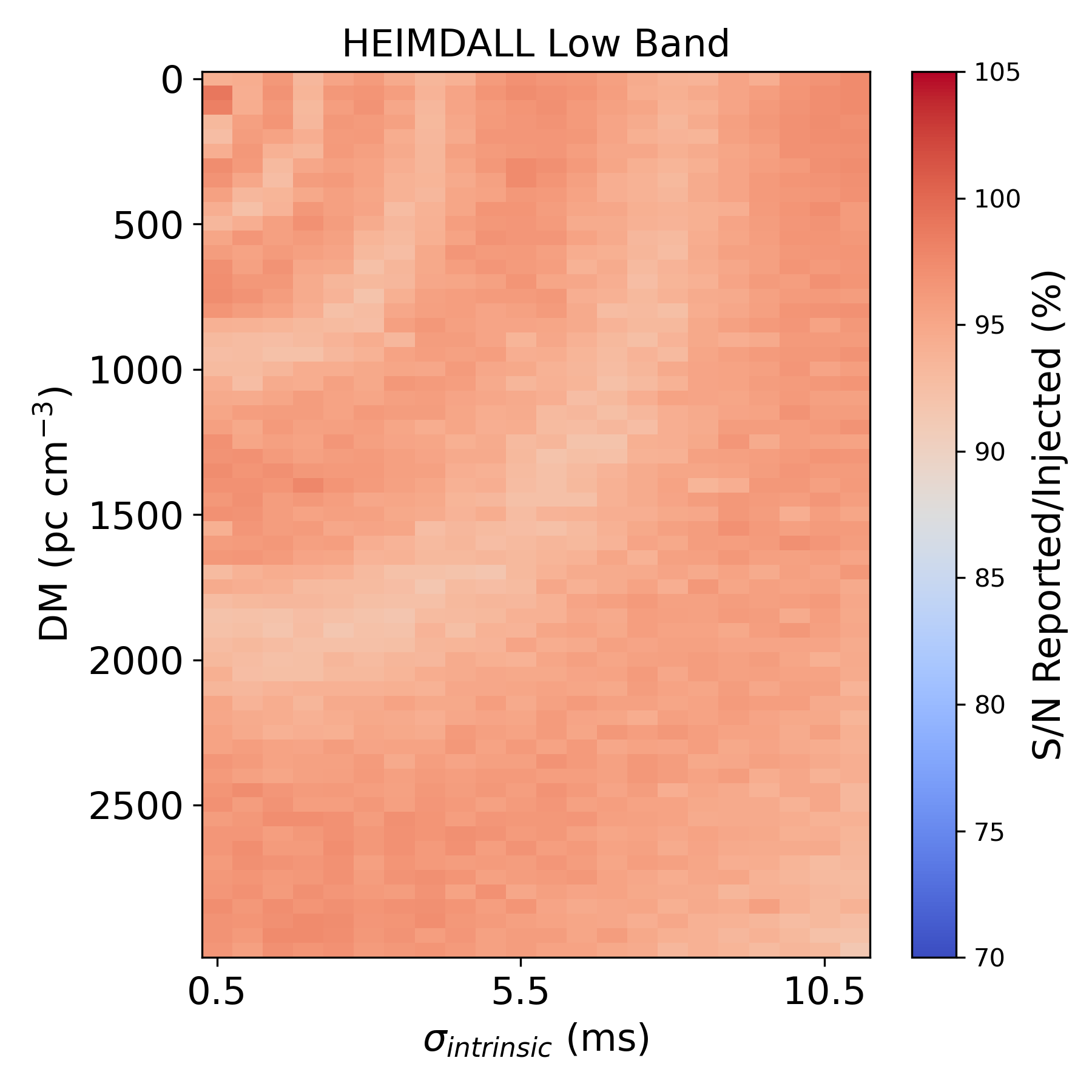}
\includegraphics[trim={0 0.0cm 0 0.0cm },clip,width=0.8\columnwidth]{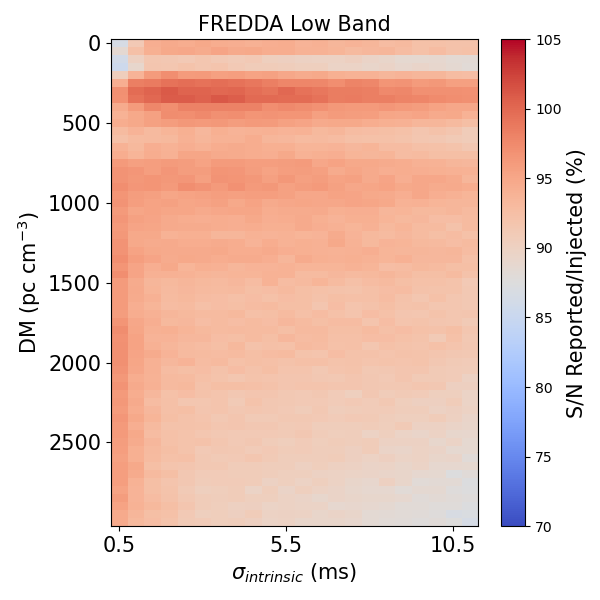}
    \caption{FREDDA and HEIMDALL reported/injected S/N of single pulses in the high and low frequency band.}
    \label{fig:single_results}
\end{figure*}

We first process the pulses in their original dispersed form.
The average S/N values reported over our $50$ iterations as a function of $\sigma_{\rm intrinsic}$ and DM are shown in Figure 
\ref{fig:single_results}. 
The results confirm that {\sc fredda} maintains a high S/N recovery across the entire parameter range.
This is especially important for high DM searches where a drop in S/N is expected due to broadening caused by DM smearing. The low SN at higher DM and widths of FREDDA low band are caused by the maximum limit of 32 sample boxcars. The scalloping effect can be seen across the parameter space for {\sc heimdall} in Figure \ref{fig:single_results}. The response is the one shown in Figure~\ref{fig:heimdall_response}, but with the peaks/troughs of the scallops at constant observed pulse width, i.e. ever lower $\sigma_{\rm i}$ as DM increases. 

\begin{figure*}
\centering
\includegraphics[trim={0 0.0cm 0 0.0cm },clip,width=0.8\columnwidth]{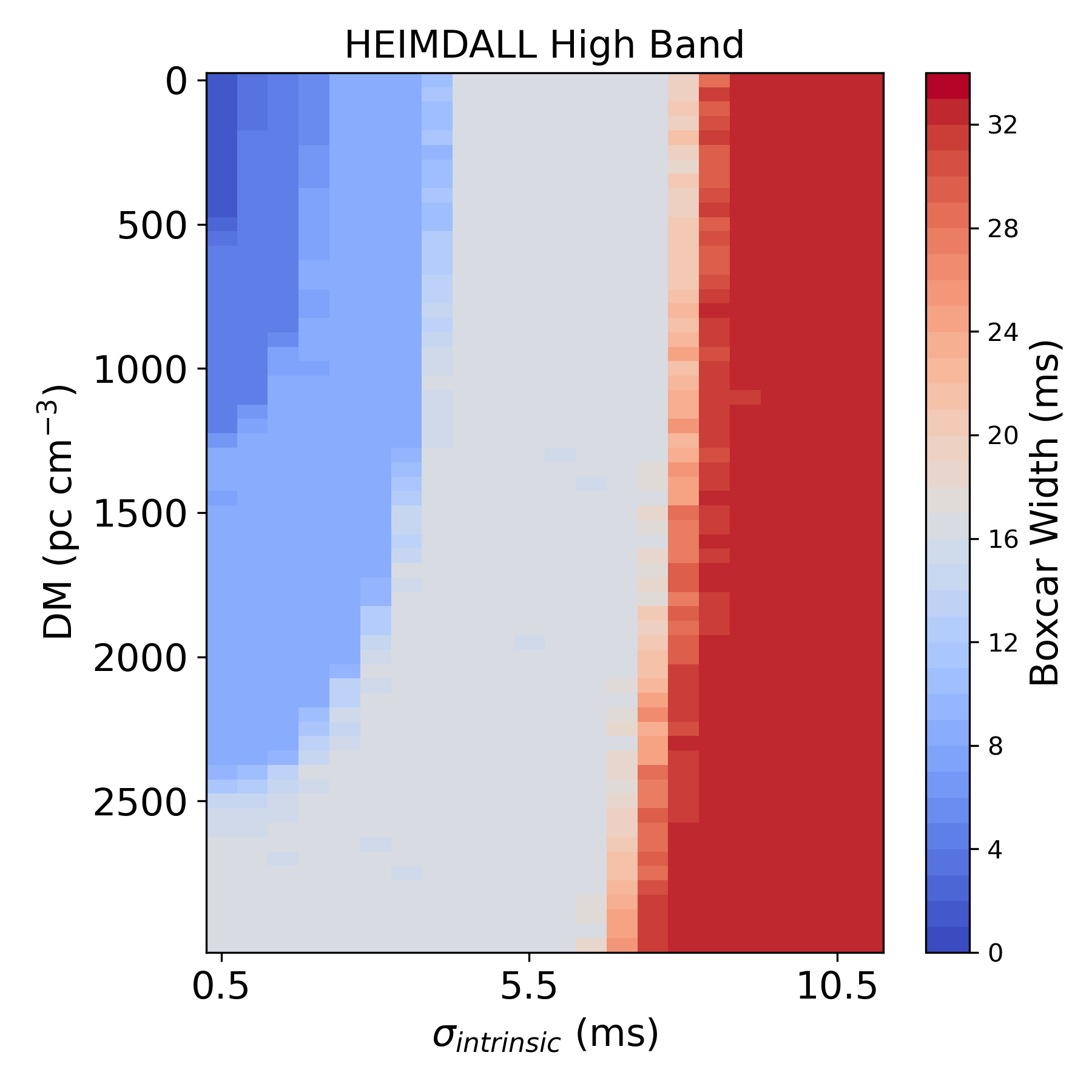}
\includegraphics[trim={0 0.0cm 0 0.0cm },clip,width=0.8\columnwidth]{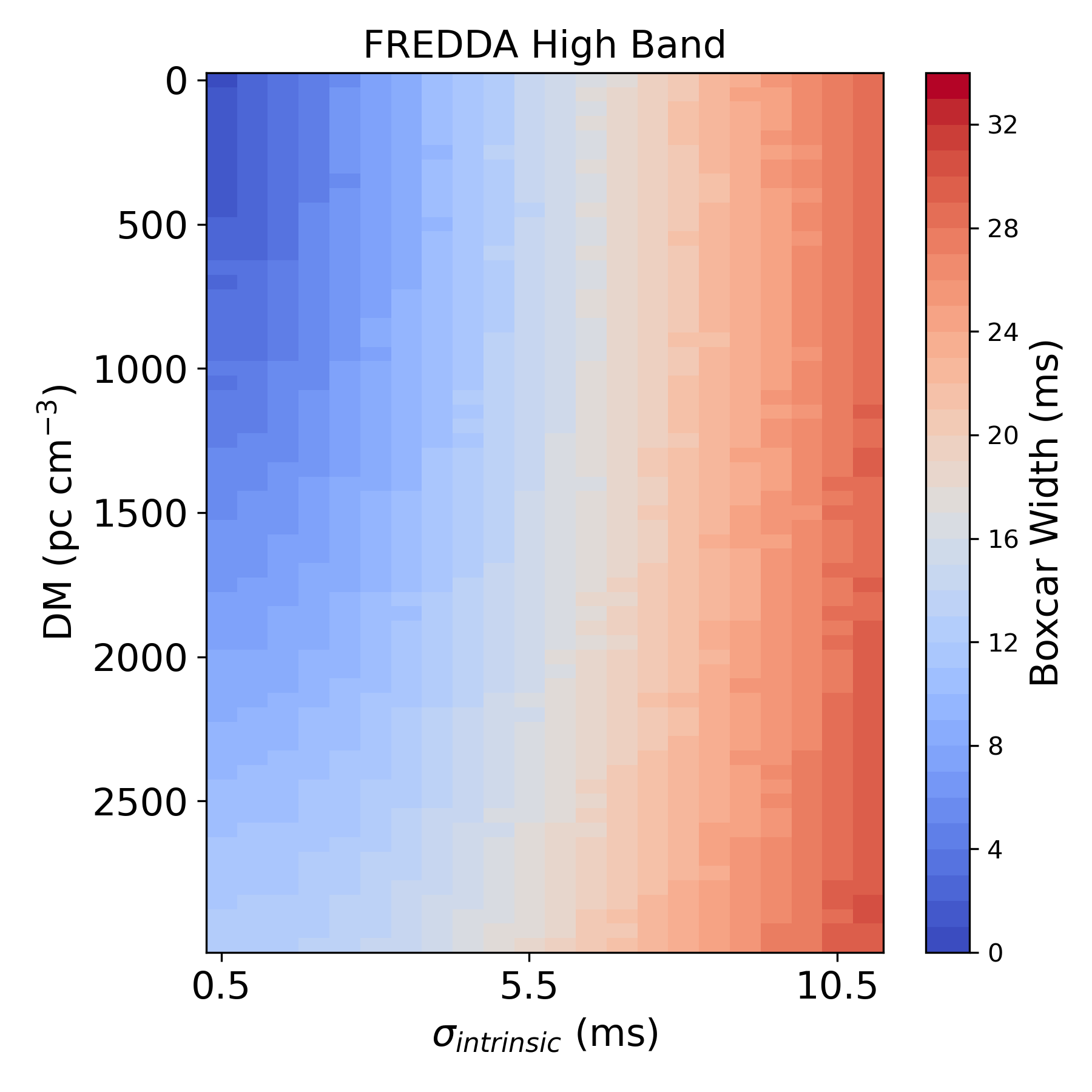}\\
\includegraphics[trim={0 0.0cm 0 0.0cm },clip,width=0.8\columnwidth]{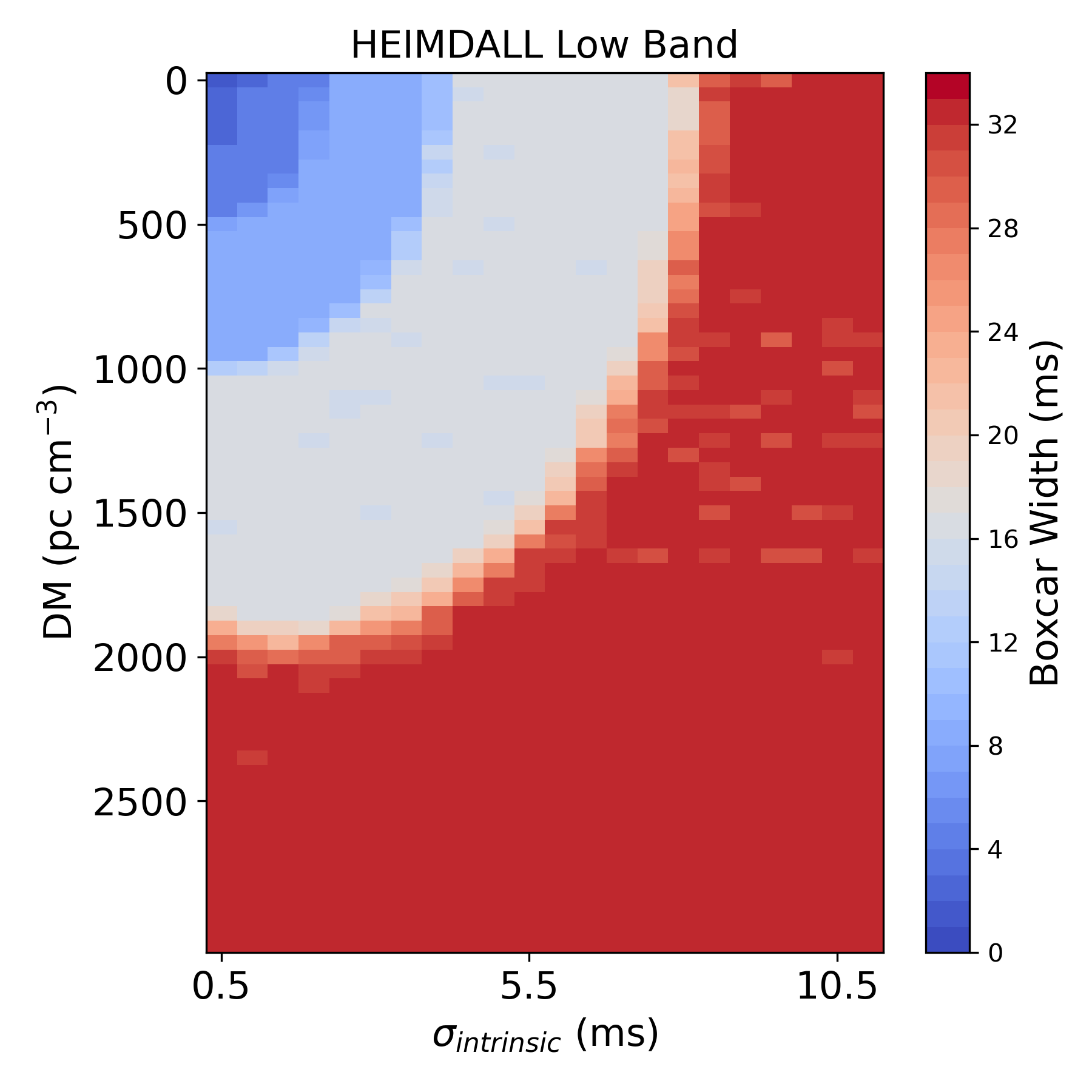}
\includegraphics[trim={0 0.0cm 0 0.0cm },clip,width=0.8\columnwidth]{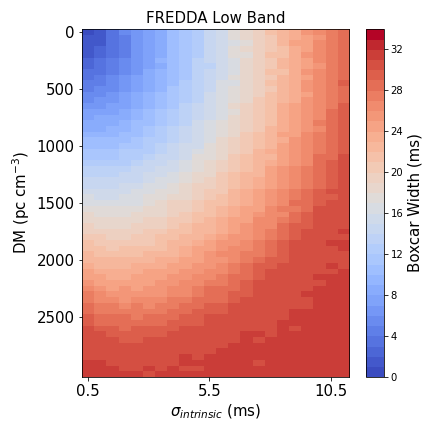}
    \caption{Average reported boxcars of injected single pulses in the high and low frequency band from FREDDA and HEIMDALL. FREDDA benefits from consecutive boxcar filters and provides a very smooth boxcar measurement.}
    \label{fig:single_boxcar}
\end{figure*}

The average boxcar values reported over $50$ iterations are displayed in Figure \ref{fig:single_boxcar}. This shows that {\sc heimdall}, as expected, is using less steps of boxcars compared to {\sc fredda}.
The total pulse width which consists of a contribution from both the intrinsic pulse width and the smearing width (which correlates with DM). This produces the ripple-like S/N response pattern of {\sc heimdall} as it applies the same fixed boxcars over wide width ranges.

\subsection{Zero-DM pulses}

\begin{figure*}
\centering
\includegraphics[trim={0 0.0cm 0 0.0cm },clip,width=0.8\columnwidth]{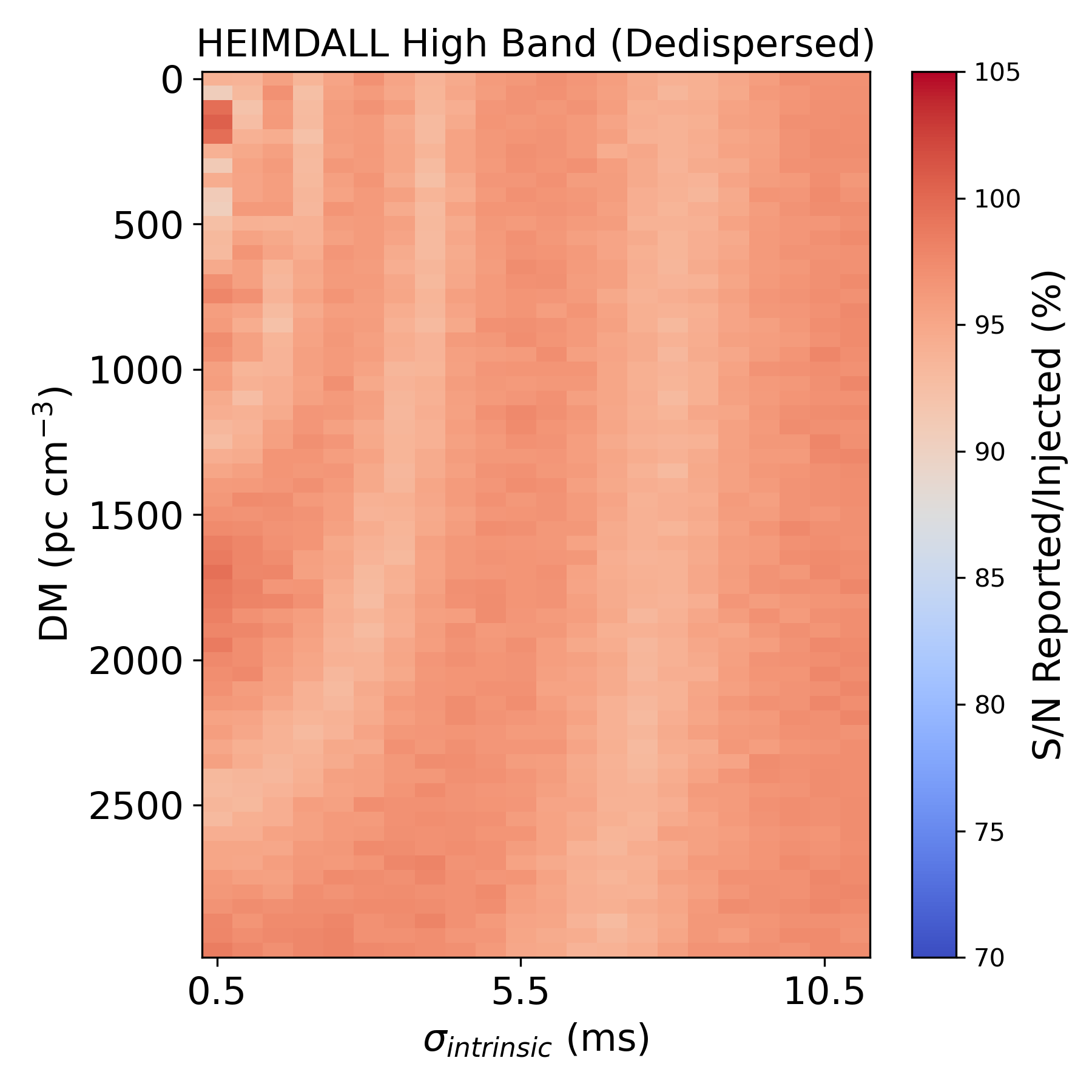}
\includegraphics[trim={0 0.0cm 0 0.0cm },clip,width=0.8\columnwidth]{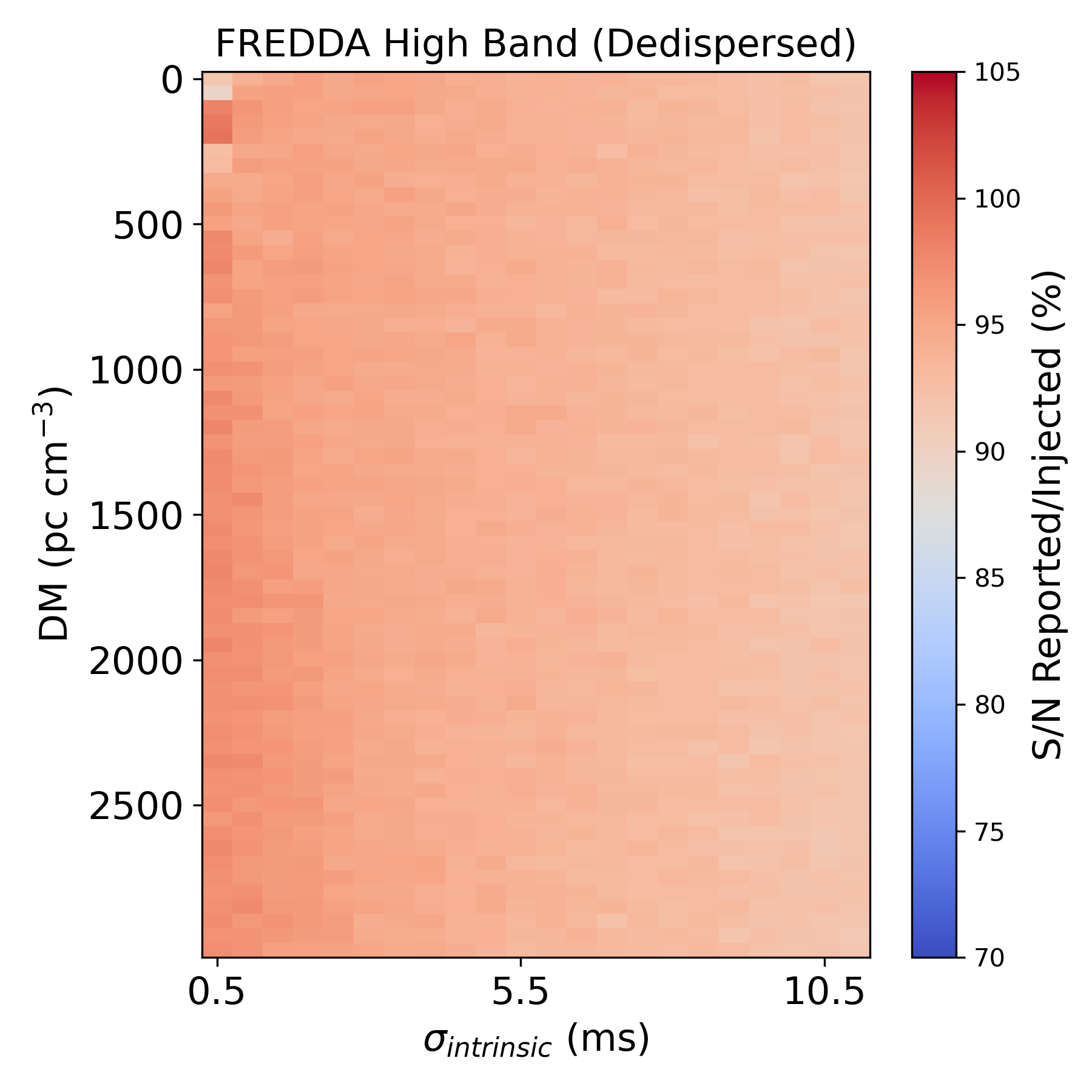} \\
\includegraphics[trim={0 0.0cm 0 0.0cm },clip,width=0.8\columnwidth]{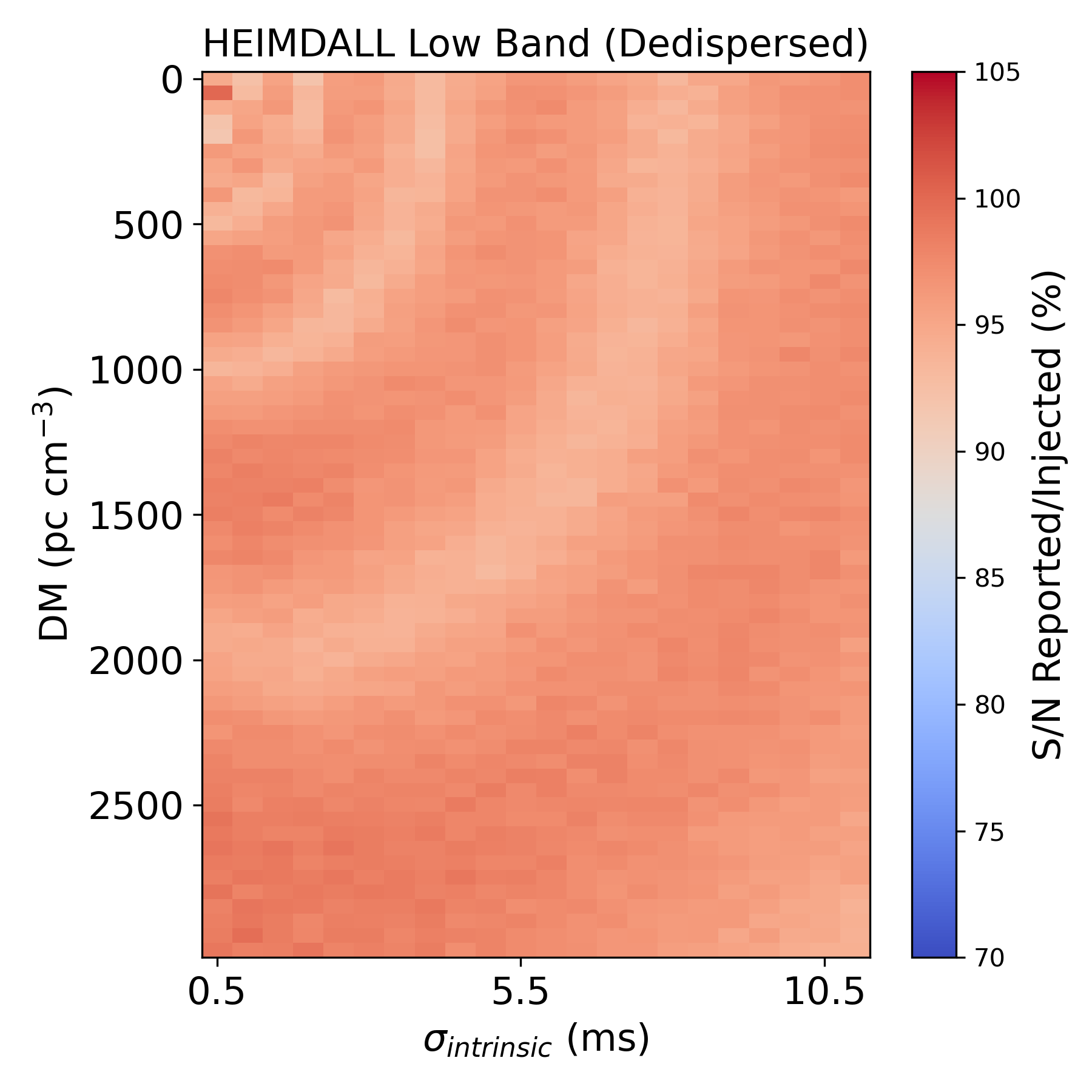}
\includegraphics[trim={0 0.0cm 0 0.0cm },clip,width=0.8\columnwidth]{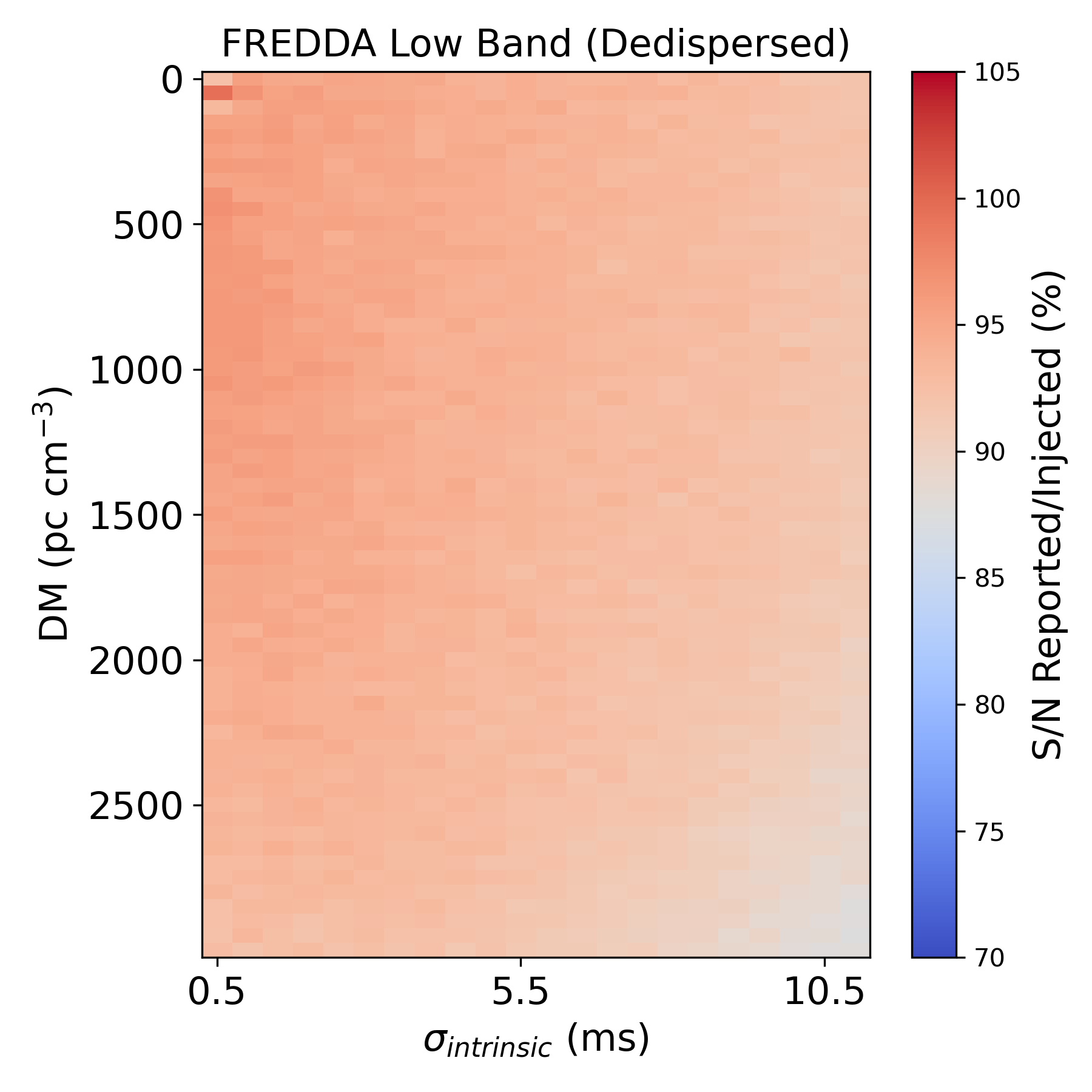}
    \caption{The S/N reported/injected by HEIMDALL and FREDDA from the perfectly incoherently dedispersed dataset.}
    \label{fig:0dm_results}
\end{figure*}

For comparison, we have a dataset that is incoherently dedispersed to the exact DM (intrachannel smearing remains). The results of processing these data are shown in Figure \ref{fig:0dm_results}.
The results are a direct examination of the S/N boxcar algorithm applied by the software when the pulse is correctly dedispersed but smeared, hence this is the theoretical best S/N achievable for each pipeline.
It can be seen that the response is consistent to the simulated estimates in Figure \ref{fig:heimdall_response} and Figure \ref{fig:fredda_response} as DM and $\sigma_{\rm intrinsic}$ correlates with the pulse FWHM.


\begin{figure*}
\centering
\includegraphics[trim={0 0.0cm 0 0.0cm },clip,width=0.8\columnwidth]{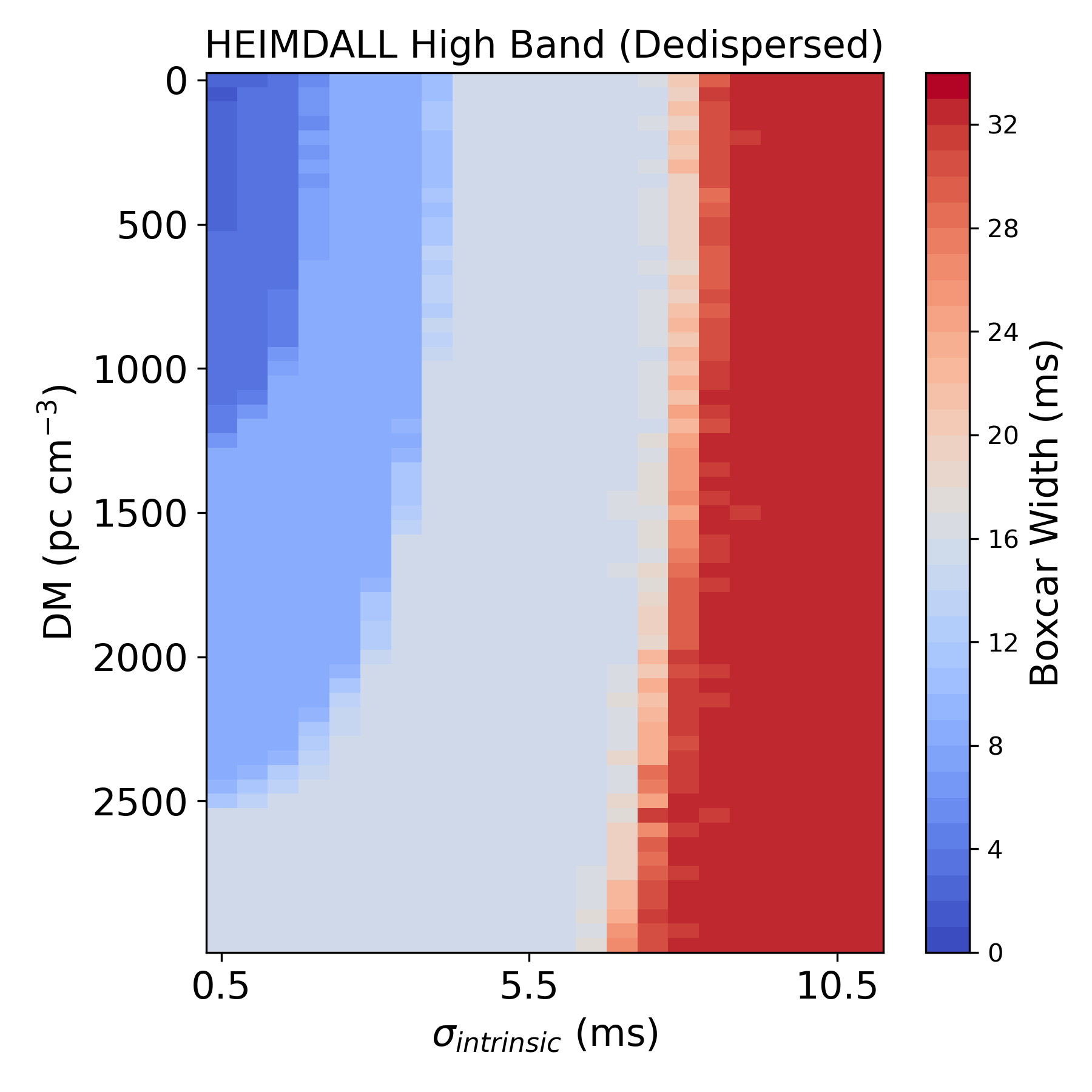}
\includegraphics[trim={0 0.0cm 0 0.0cm },clip,width=0.8\columnwidth]{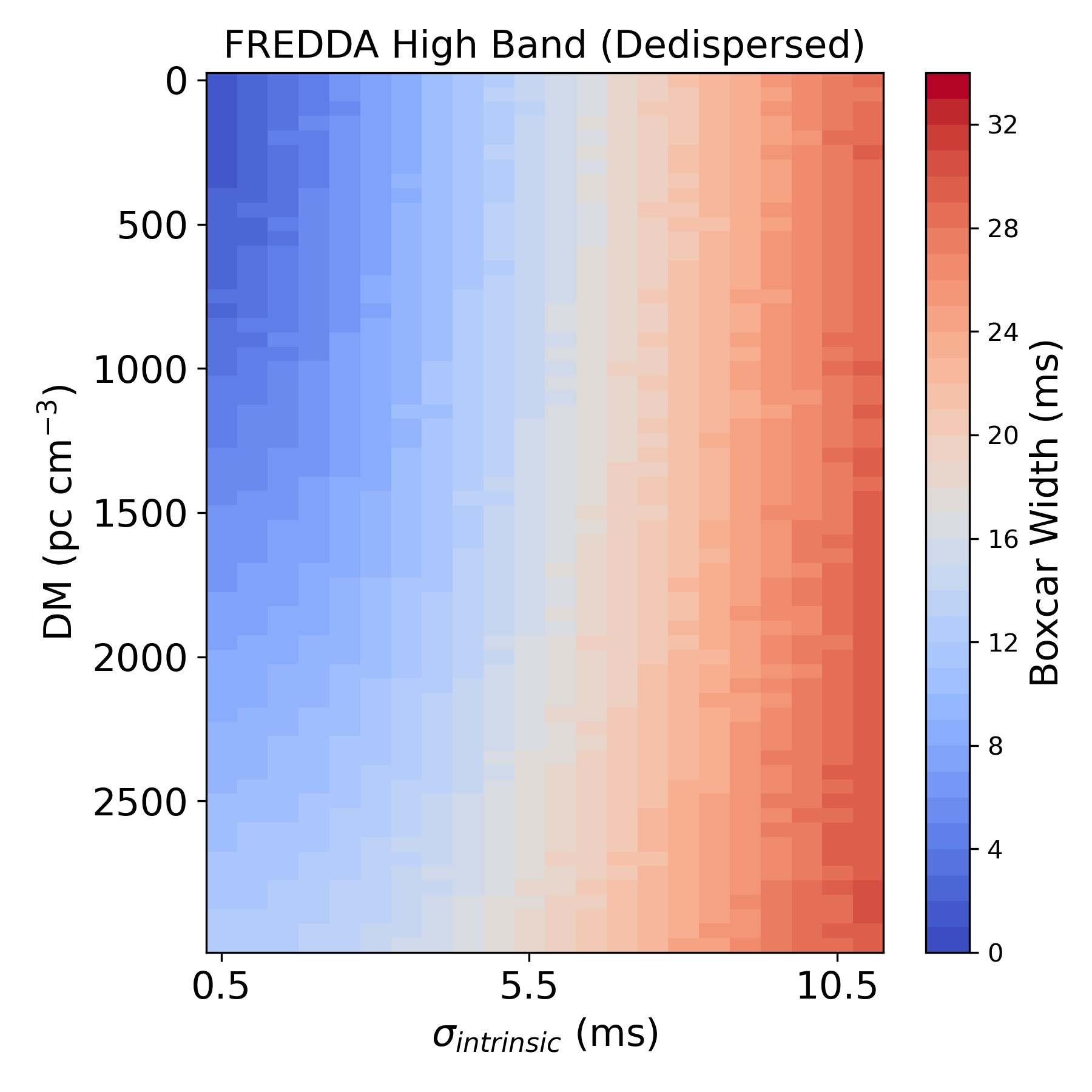}\\
\includegraphics[trim={0 0.0cm 0 0.0cm },clip,width=0.8\columnwidth]{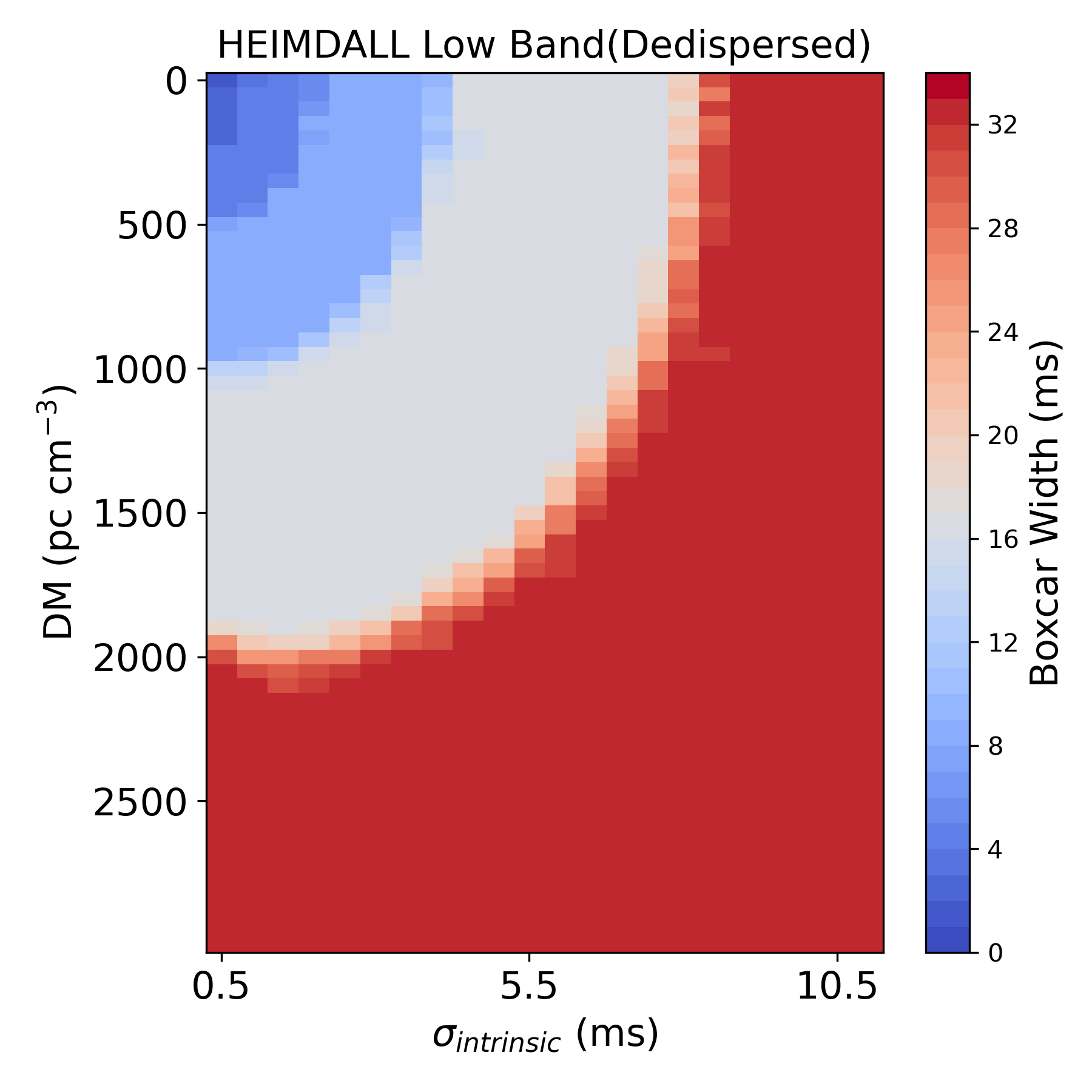}
\includegraphics[trim={0 0.0cm 0 0.0cm },clip,width=0.8\columnwidth]{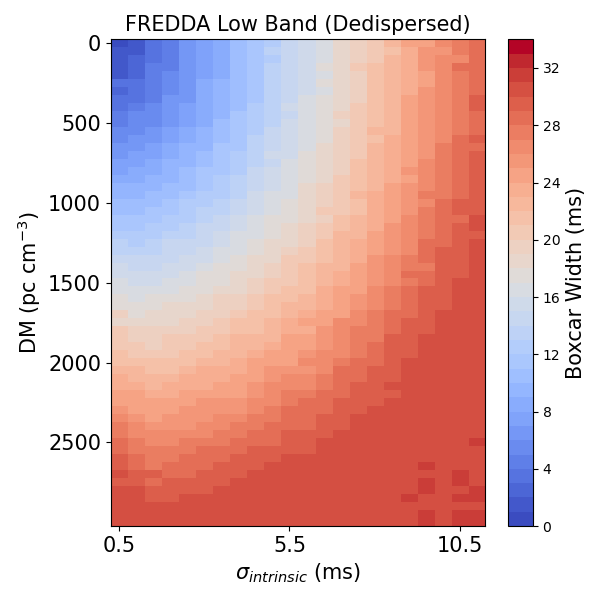}
    \caption{Reported boxcar from measuring dedispersed pulses with HEIMDALL and FREDDA.}
    \label{fig:boxcar_0dm}
\end{figure*}


\subsection{FREDDA high S/N at intermediate DM}

We find that for some parts of the parameter space the S/N reported by FREDDA from dispersed pulses is nominally better than the theoretical maximum, and even exceeds 100\% of that injected.
For example, in the high frequency range, the maximum S/N reported by {\sc fredda} at around $\rm{DM\sim 750\ pc\ cm^{-3}}$
reaches 102\% of the injected S/N. 
For the low frequency dataset, the S/N bump shifts to a lower range at $\rm{DM\sim 350\ pc\ cm^{-3}}$ and is less emphatic.
This result greatly affects the statistics of FRBs discovered by \textsc{fredda}. For these data there are no islands in the parameter space where excessively high S/N values are obtained.

\begin{figure}
\centering
\includegraphics[trim={0 0.0cm 0 0.0cm },width=0.8\columnwidth]{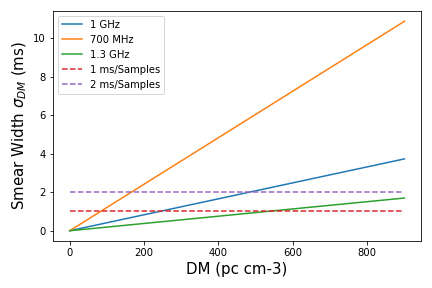}

    \caption{Dispersion smearing at different frequencies in relation to DM of the pulse. We also draw two horizontal lines identifying width of 1~ms and 2~ms, which corresponds to 1-2 time samples in this work. We identify that the smearing width is between 1-2 time samples at 200-400 \dm at 1~GHz and 600-1000 \dm at 1.3~GHz. This DM range corresponds to the unusual S/N rise in FREDDA}
    \label{fig:minimal_smear}
\end{figure}

We deem it to be important that this issue only appears for dispersed pulses, i.e. not for the perfectly dedispersed pulses. This would indicate that the underlying issue is related to the dedispersion process, not the pulse search part of the algorithm.
We note that the channel smearing width at the top of the band for a burst at these DM ranges in each frequency band is between $1-2$~ms (corresponding to $1-2$ time sample in this work) as shown in Figure \ref{fig:minimal_smear}. This indicates that the S/N reported is affected by how the algorithm collects the dispersed signal for the S/N calculation, e.g. if the samples occupied by the dispersive sweep were under-estimated so too would the rms noise, resulting in an over-estimated S/N.

\begin{figure}
    \centering
    \includegraphics[trim={0 0.0cm 0 0.0cm },width=0.8\columnwidth]{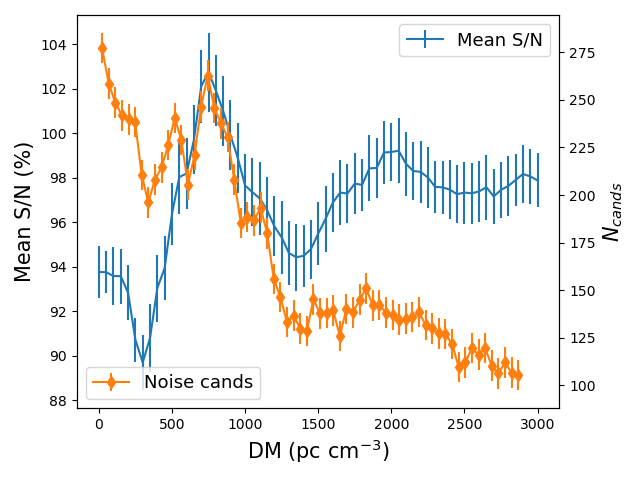}
    \caption{The number of noise candidates above 3 sigma in comparison to the average S/N reported by {\sc fredda} over DM in 60 seconds of high frequency band data. The data points of the noise candidate count are the averaged bins of 64 dm steps.}
    \label{fig:noise-dm}
\end{figure}

We also examine how {\sc fredda} calculates the noise across different DMs by processing 20 seconds of Gaussian white noise data under the same settings. We show in Figure \ref{fig:noise-dm} the number of noise candidates above $3\sigma$ in comparison to the average S/N in high frequency band data.

The number of noise candidates is expected to drop slightly as DM increases, as the larger DM increases the minimum width and time delay of pulses which reduces the effective search length of the data, which will be more significant in shorter length data (20 seconds).
We however see significant fluctuations in the figure: a large increase in the number of noise candidates at around $700-800$~\dm, which correlates with the rise of reported S/N in {\sc fredda}. 
We also speculate that the lower-DM S/N dips may be correlated with the dip in the number of noise candidates at 400 \dm and 1400 \dm respectively.
This indicates that there is an error with the noise estimation function that may be the cause of the S/N inconsistency.

Further investigation of the algorithm is being conducted to explain the phenomena and remove this software error in future searches. But this feature exists in all searches to date and so the response function for the ASKAP FRB sample, for instance, should include it as it is calculated here.

\subsection{Search Completeness}
The sensitivity of the pipelines are well characterised from this work. For statistical studies on ASKAP FRBs and the FRB searches conducted with {\sc heimdall}, we can utilise the reported S/N results to estimate the real S/N/fluence of these FRBs. We assess the search completeness at different frequencies for both pipelines based on the $\log N-\log S$ relation of a uniformly distributed source population in a Euclidean Universe with a slope of -1.5. 
We relate the search flux density $\rm{S_{search}}$ with the target search sensitivity $\rm{S_{target}}$
with $\rm{S_{search}=S_{target}\times {\cal R}}$.
Here ${\cal R}$ is the ratio between the reported and the injected S/N (e.g Figure \ref{fig:single_results}) scaled by the S/N down-scaling response function for constant S/N pulses (see \S~\ref{sec:rescaling} and Figure \ref{fig:downscale}). 

When the reported S/N is higher than 100\% of that injected, the recovery is considered as 100\% (e.g parts of the {\sc fredda} results), as it has reached target sensitivity. The real flux density of the population actually detected is lower than the target sensitivity of the observations.
The search completeness is therefore expressed in the terms of the cumulative $\rm{N_{search}/N_{target}}$.

Our calculations are shown in Table \ref{tab:logn-logs}. 
The reported S/N maps show that pipelines are only probing $\sim 90\%$ of the designated parameter space in this work due to not reaching the real target sensitivity.
At detection thresholds or for real time searches, the search completeness will be further lowered, according to the noise statistics. 

\begin{table}
    \centering
    \begin{tabular}{l|ll|c}
         Pipeline& Band & dm\_tol& $\rm{N_{search}/N_{target}}$\\
         \hline
         \multirow{2}{*}{{\sc fredda}}&High&-- &$92.6\%$  \\ 
         &Low & -- &$87.5\%$ \\
         \hline
         \multirow{2}{*}{{\sc heimdall}}&High
         & 1.01
         & $90.2\%$\\
         &Low&1.01&$90.0\%$ \\
         \hline  
                  \multirow{2}{*}{{\sc heimdall}}&High&1.25& $86.7\%$\\
         &Low&1.25&$85.1\%$ \\
         \hline  
         \end{tabular}
    \caption{Search completeness compared to target sensitivity for both pipelines in high and low frequency bands. {\sc heimdall} shows consistency over frequency, while {\sc fredda} performs better at high frequency.}
    \label{tab:logn-logs}
\end{table}


\section{Conclusion}
In this work, we utilise simulated single pulses to test the performance of the FRB search pipeline {\sc fredda} and {\sc heimdall}. 
We simulated pulses in Gaussian white noise with no interference across a parameter space of DM < 3000 \dm and intrinsic pulse standard deviation below < 11 ms.

Our results show that both {\sc heimdall} and {\sc fredda} perform effectively in detecting single dispersed pulses under ideal conditions. When a very fine DM tolerance of \textsc{heimdall} is taken, it matches the performance of \textsc{fredda}, with both pipelines sensitive to $\sim 90\%$ of the search volume. 
However, 
in exchange for computation cost and low latency, if one reduces the DM tolerance for {\sc heimdall}(as is typically done) the performance is worse at $\sim 86\%$.
We identify an issue in  {\sc fredda} where the reported S/N is higher than 100\% of that injected likely due to dispersion smearing dominating pulse width.


We reconfirm results from previous tests \citep{2015MNRAS.447.2852K} that pipelines using a non-consecutive set of boxcar match filters such as {\sc heimdall} will receive a signal to noise penalty on pulses in between the boxcar widths. 
We further broadened the response calculation to include DM and a number of other additional subtle effects. 
We demonstrate that using consecutive boxcars (e.g {\sc fredda}) is more sensitive to search for pulses in coarse time resolution real-time data. 

We also calculate a theoretical search completeness using the S/N response distribution over the DM and pulse width parameter space obtained in this work for {\sc heimdall} and {\sc fredda}. 
We show that {\sc fredda} achieves a search completness of 92.6\% at 1.1-1.4 GHz.
Depending on the settings the search completeness of {\sc heimdall} in that frequency band is between 86.7\% to 90.2\%.

We believe this result demonstrates the importance of understanding the pipeline search completeness not only for intial detection but also for FRB population and statistical studies.
It is essential to the FRB community that current and future large FRB surveys such as CHIME/FRB and DSA-2000 to investigate such underlying effects. The data and software for this work has been made publicly available for further simulation and analysis use on other pipelines.


\section*{Acknowledgements}
HQ thanks David McKenna, Vivek Gupta, Wael Farah, Andrew Jameson and Adam Deller for useful comments on FREDDA and HEIMDALL for this work.
HQ and EFK acknowledge support from Fondation MERAC (Project: SUPERHeRO, PI: Keane) and are grateful to Aaron Golden for support when working in Galway. This work was performed on the OzSTAR national facility at Swinburne University of Technology and the SKA Observatory Science Computing facilities. The OzSTAR program receives funding in part from the Astronomy National Collaborative Research Infrastructure Strategy (NCRIS) allocation provided by the Australian Government.


\section*{Data Availability}

The source code to simulate FRBs for this work is publicly available at the following repository:
\url{https://github.com/hqiu-nju/simfred/}.
The data results in this work, supplement material and script to reproduce the simulations are available in the following repository : \url{https://github.com/hqiu-nju/CRAFT-ICS-pipeline}



\bibliographystyle{mnras}
\bibliography{main} 




\appendix
\section{S/N Injection}
\begin{figure}
    \centering
\includegraphics[width=0.95\columnwidth]{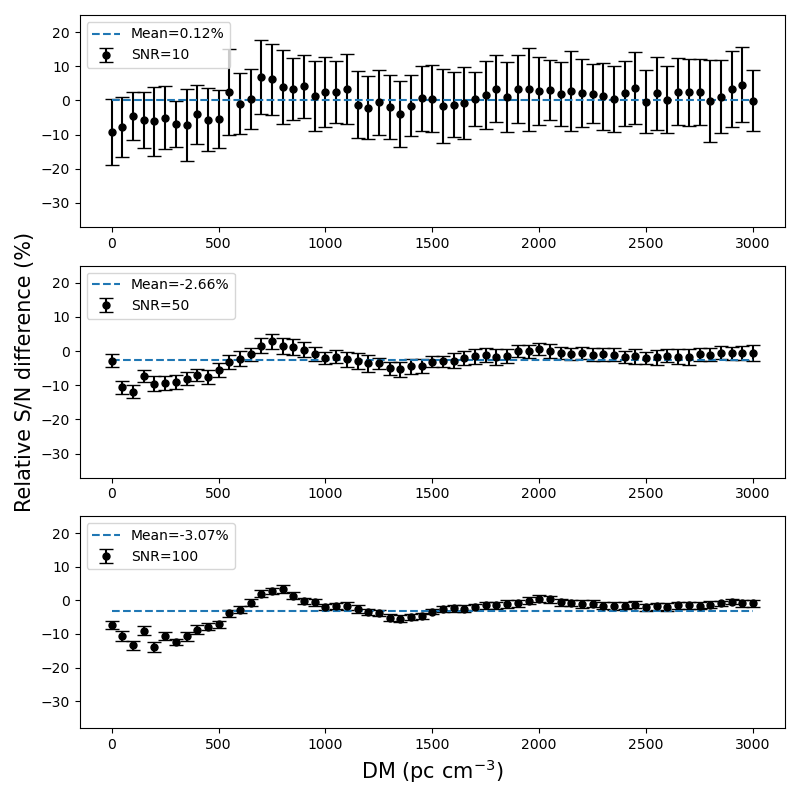}
    \caption{The relative difference of reported S/N of narrow pulses at different dispersion measures with an S/N of 10, 50 and 100. The mean S/N loss of each dataset is plotted as the blue dash line for comparison.}
    \label{fig:snr_50}   
\end{figure}

We present the results of injecting narrow ($\sigma=0.5$ ms) pulses across DM $=0-3000$~\dm and S/N $=10-100$.
Figure \ref{fig:snr_50} shows that the response is the same regardless of the input S/N except for a slight loss as the input S/N gets very high.
This latter effect is understood; it is the due to the very high S/N pulses contaminating the background noise calculation in {\sc fredda}. 
We note that scallop effects also play a role in the relative S/N loss. The full dataset is available in the Appendix A directory of the repository.
\label{sec:snr}
\section{Scattering}
\label{sec:scattering}
We also designed a simulation with scatter broadened pulses to test the pipeline capability with scattered and other asymmetric pulse profiles.
The scatter broadening of pulses may affect the S/N as predicted, 
as the pulse duration is extended leading to a wider boxcar.

We do not find strong evidence of differences in S/N between the scattered pulses and Gaussian single pulses. 
The average S/N of scattered pulses varies similarly to Gaussian pulses as scattering time, DM and intrinsic width both affect the pulse duration. 
A longer pulse duration caused by scatter broadening leads to larger boxcars resulting in the similar S/N decrease with pulse width 
However, the correlation between DM and S/N is much more significant. 

The simulation code is available in the repository to regenerate this dataset for further studies.

\section{Fractional Bandwidth}
\label{sec:fractionalband}

Many FRBs have been observed to exhibit frequency profiles concentrated into a narrow bandwidth \citep{Kumar2021,2021ApJ...923....1P}. Additionally, narrowband radio frequency interference (RFI) or interstellar scintillation may reduce the effective bandwidth.
Such effects may interact with the time-frequency templates used by FREDDA to search for dispersed pulses, affecting the S/N response.

We choose to simulate an extreme example of such spectral effects so as to gauge the magnitude of the impact on the algorithm. We simulated a dataset of Gaussian pulses with the same parameter settings used in the main section of this paper but only with $1/6$ the bandwidth; this is comparable to the 65\,MHz bandwidth observed from FRB 20190711A \citet{Kumar2021}. We choose band occupancy to be concentrated into the third sextet of the band. Pulses were scaled to S/N=20 to prevent saturation of the samples in our 8-bit data.
\begin{figure}
    \centering
    \includegraphics[width=0.95\columnwidth]{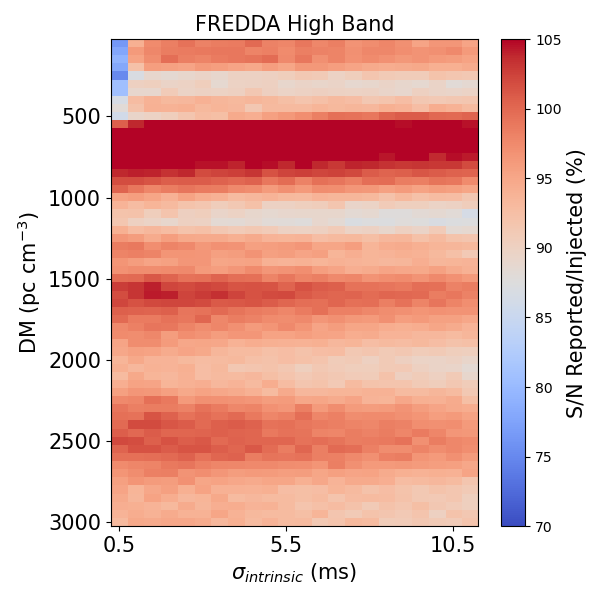}
    \caption{FREDDA reported/injected S/N of fractional bandwidth pulses in high frequency band.}
    \label{fig:fractionalbands}
\end{figure}

We show in Figure \ref{fig:fractionalbands} the result from {\sc fredda}, where the overall response remains consistent with that of the overall performance from the broadband pulses in Figure~\ref{fig:single_results}. The scalloping effect across dispersion measure is also observed, but is more pronounced. 
This suggests that the DM scallop structure is present over the entire bandwidth, but its effects are washed out in the case of broadband pulses at high DM, due to different scalloping as a function to frequency. 



\bsp	
\label{lastpage}
\end{document}